\documentclass[aps,pra,twocolumn,superscriptaddress,footinbib]{revtex4-2}
\usepackage{mathtools}
\usepackage{amssymb}
\usepackage{amsmath,amssymb,bm}
\usepackage{mathrsfs}
\usepackage{graphicx}
\usepackage{epstopdf}
\usepackage{latexsym}
\usepackage[normalem]{ulem}
\usepackage[usenames,dvipsnames]{color}
\usepackage{hyperref}
\usepackage{natbib}
\usepackage{comment}
\usepackage[dvipsnames]{xcolor}

\newcommand{\ic}{i}

\renewcommand{\vec}[1]{\boldsymbol{\mathbf{#1}}}
\DeclarePairedDelimiter\abs{\lvert}{\rvert}
\newcommand{\Id}{\mathbb{I}}
\newcommand*{\trans}{T}
\renewcommand{\Re}{\operatorname{Re}}
\renewcommand{\Im}{\operatorname{Im}}

\newcommand{\vpsi}{\vec{\Psi}}
\newcommand{\dvpsi}{\delta\vec{\Psi}}
\newcommand{\vl}{\vec{\Lambda}}
\newcommand{\vls}{\vec{\Lambda}_\parallel}
\newcommand{\vlp}{\vec{\Lambda}_\perp}
\renewcommand{\vr}{\vec{r}}
\newcommand{\tvl}{\tilde{\vec{\Lambda}}}
\newcommand{\tvls}{\tilde{\vec{\Lambda}}_\parallel}
\newcommand{\tvlp}{\tilde{\vec{\Lambda}}_\perp}
\newcommand{\tvu}{\tilde{\vec{u}}}
\newcommand{\tvv}{\tilde{\vec{v}}}
\newcommand{\vs}{\vec{s}}
\newcommand{\tvs}{\tilde{\vec{s}}}
\newcommand{\vF}{\vec{F}}
\newcommand{\Hscal}{H_\text{scalar}}
\newcommand{\Hs}{H_s}
\newcommand{\cH}{\mathcal{H}}
\newcommand{\cHsgp}{\vec{\mathcal{H}}_\text{SGP}}
\newcommand{\cHscal}{\mathcal{H}_\text{scalar}}
\newcommand{\cHs}{\vec{\mathcal{H}}_s}
\newcommand{\vhs}{\vec{h}_s^\text{SMA}}
\newcommand{\vUsma}{\vec{U}_s^\text{SMA}}
\newcommand{\mun}{\mu_n}
\newcommand{\mus}{\mu_s}
\newcommand{\vSl}{\vec{\mathcal{S}}}

\newcommand{\cLgp}{\vec{\mathcal{L}}_\text{GP}}
\newcommand{\cLscal}{\vec{\mathcal{L}}_\text{scalar}}
\newcommand{\cLs}{\vec{\mathcal{L}}_s}
\newcommand{\tcLgp}{\tilde{\vec{\mathcal{L}}}_\text{GP}}
\newcommand{\tcLscal}{\tilde{\vec{\mathcal{L}}}_\text{scalar}}
\newcommand{\tcLs}{\tilde{\vec{\mathcal{L}}}_s}
\newcommand{\tcLpar}{\tilde{\mathcal{L}}_\parallel}
\newcommand{\tcLperp}{\tilde{\mathcal{L}}_\perp}
\newcommand{\tvSl}{\tilde{\vec{\mathcal{S}}}}
\newcommand{\cQ}{Q}
\newcommand{\cQs}{\vec{Q}_s}
\newcommand{\modeperp}[1]{u_{\perp#1}}
\newcommand{\modeppar}[1]{u_{\parallel#1}}
\newcommand{\modehpar}[1]{v_{\parallel#1}}
\newcommand{\cM}{\vec{\mathcal{M}}}
\newcommand{\cC}{\vec{\mathcal{C}}}
\newcommand{\cPperp}{\vec{\mathcal{P}}_\perp}
\newcommand{\cPpar}{\vec{\mathcal{P}}_\parallel}

\begin{document}

\title{Dynamics of spinor Bose-Einstein condensates close to spin-spatial
resonances}
\author{W. Wills}
\affiliation{Department of Physics and Astronomy, Washington State University,
Pullman, WA 99164-2814, USA}
\author{D. Blume}
\affiliation{Homer L. Dodge Department of Physics and Astronomy, The University
of Oklahoma, Norman, Oklahoma 73019, USA}
\affiliation{Center for Quantum Research and Technology, The University of
Oklahoma, Norman, Oklahoma 73019, USA}
\author{Q. Guan}
\affiliation{Department of Physics and Astronomy, Washington State University,
Pullman, WA 99164-2814, USA}
\date{\today}

\begin{abstract}
We develop a coupled-channel framework to describe the dynamics of spinor
Bose-Einstein condensates (BECs), with particular emphasis on the behavior near
resonances between spin dynamics and spatial excitations. Taking advantage of
the disparity between the spin-dependent and spin-independent scattering
lengths in typical spinor BECs, the Bogoliubov modes of the spin-independent
part of the full system Hamiltonian provide an efficient set of basis functions
for describing the system dynamics in a coupled-channel framework. For
quadratic Zeeman shifts far from any resonance, the system can be described by
a single spatial wavefunction during the spin dynamics, i.e., the so-called
single-mode approximation holds. By tuning the quadratic Zeeman shift, we find
resonant excitations of the Bogoliubov modes, which can be classified into two
categories: those with particle–hole correlations and those without
particle-hole correlations. We show that the beyond-quadratic-order terms that
are neglected in standard Bogoliubov theories become increasingly important for
capturing the long-time dynamics of the system near resonances. The
coupled-channel framework is benchmarked against results from 1D
Gross–Pitaevskii equation simulations. The framework developed in this work not
only provides a numerically efficient tool for describing spinor BEC dynamics
governed by different length scales, but also provides a clean physical
interpretation of resonance phenomena in spinor BECs. Applications of this
approach to other systems and extensions to the beyond-mean-field regime are
also discussed.
\end{abstract}

\maketitle

\section{Introduction}

Spinor Bose–Einstein condensates (BECs), i.e., BECs with multiple internal spin
degrees of freedom, are unique platforms due to the presence of intrinsic
elastic and inelastic spin collisions, which, for example, are crucial for
generating spin entanglement and realizing SU(1,1)
interferometers~\cite{KAWAGUCHI2012253, Dan2023RevModPhys}. Spinor BECs are
frequently realized in $^{23}$Na and $^{87}$Rb quantum
gases~\cite{Dan2023RevModPhys}, for which the spin-dependent scattering length
is much smaller than the spin-independent one. The disparity between the two
associated energy scales enables the system to serve as a clean platform for
studying pure spin physics in situations where the spatial part of the
wavefunction remains essentially frozen, even when the spin degrees of freedom
are in highly non-equilibrium states~\cite{You_PhysRevA2002}. Within the
single-mode approximation (SMA), the spin dynamics are governed by the
spin-interaction energy and the quadratic Zeeman
shift~\cite{Ho_PhysRevLett1998, Ho_PhysRevA2000, doi:10.1143/JPSJ.67.1822,
You_PhysRevA2005}. With magnetic fields and microwave
dressing~\cite{Bloch_PhysRevA2006, Liu_PhysRevA2014}, the latter can be tuned
over a wide range, which further enhances the system’s tunability and enables
the study of diverse quantum phenomena, including equilibrium and
non-equilibrium quantum phase transitions~\cite{Sadler2006,
PhysRevLett.107.195306, PhysRevLett.102.125301, Gessner_PhysRevLett2021,
Duan_PhysRevA2019, Duan_PhysRevA2018, Zhang_PhysRevResearch2023, Liu_arXiv2025,
Duan_PhysRevLett2020, Guan_PRA2025}, quantum state
engineering~\cite{doi:10.1126/science.abd8206, YouLi_Science2017, Mao_PRL2023},
and quantum metrology~\cite{PhysRevA.98.023620, Mao_PRL2023,
Oberthaler_PhysRevLett2016, doi:10.1126/science.1208798,
Smerzi_PhysRevLett2015}.

Recent work by two of us and collaborators shows that the SMA can be extended
to scenarios where the spin components share a single common time-dependent (as
opposed to time-independent) spatial wavefunction that may undergo surprisingly
strong dynamics over timescales that are long compared to the coherent spin
evolution~\cite{Guan_SpinordSMA_PRA2023}. In this dynamical single-mode
approximation (dSMA), the spin and spatial degrees of freedom are
asymmetrically coupled: the spin dynamics depend on the spatial part of the
wavefunction, while the spatial part of the wavefunction remains largely
independent of the spin dynamics. Exploiting the dynamical spatial wavefunction
as an additional control knob, we used the $^{23}$Na spinor BEC as a quantum
simulator to study Landau–Zener tunneling during Bloch
oscillations~\cite{Guan_SpinorTunneling_PRA2023} and to engineer the dynamical
phase diagram under periodic modulation of the trap
frequency~\cite{Guan_SpinorDPT_PRA2024}.

Similar to the SMA, the validity of the dSMA stems from the energy scale
separation between the spin-dependent and spin-independent parts of the spinor
BEC Hamiltonian. A natural follow-up question is whether the spatial dynamics
can be affected by the non-equilibrium nature of the spin degrees of freedom.
Indeed, this occurs in multiple scenarios. An iconic example is the formation
of spin domains induced by quenching a parameter of the spinor BEC across a
quantum phase transition, where the condensate size is much larger than the
characteristic spin length scales, such as the spin healing length and the
length scale associated with the quadratic Zeeman shift~\cite{Stenger1998,
Jimenez-Garcia2019, Hirano_PhysRevA.82.033609, PhysRevLett.110.165301,
PhysRevLett.122.095301}. In such large spinor BECs, recent work has also
demonstrated that spin chaos enhances the turbulent behavior of the spatial
dynamics~\cite{PhysRevResearch.6.L032030,
lee2026enhancementdampingturbulentatomic}. For trapped spinor BECs whose size
is much smaller than the characteristic spin length scales, spatial excitations
can be, as has been confirmed experimentally, resonantly triggered by the spin
dynamics when the quadratic Zeeman shift is close to certain characteristic
values~\cite{PhysRevLett.105.135302, PhysRevA.88.053624,
PhysRevLett.104.195303, qexp}.

The mean-field Gross–Pitaevskii equation and Bogoliubov theory are powerful
theoretical tools for understanding non-equilibrium phenomena in spinor BECs,
such as dynamical instabilities~\cite{KAWAGUCHI2012253}. Several works,
including those by two of us and our collaborators, have shown that the
resonant value of the quadratic Zeeman shift for trapped spinor BECs can be
predicted using the spectrum of a particle confined in an effective potential
formed by the external trap and the background atomic density
distribution~\cite{PhysRevLett.104.195303, qth}. This paper extends previous
work by developing a Bogoliubov-theory-based framework that captures both the
resonance conditions and the long-time dynamics near such resonances. Going
beyond the effective potential picture, which corresponds to excitations with
no particle–hole correlations, our theory also predicts a new class of
resonances that exhibit particle-hole correlations. By studying the long-time
dynamics of the system near resonances, we identify the contributions of
higher-order terms, which are neglected in standard Bogoliubov theory and
represent interactions between quasi-particles. These effective
excitation-excitation interactions have previously been found to be important
for describing different damping mechanisms such as Beliaev damping in spinor
BECs~\cite{PHUC2013158}.

In general, Bogoliubov theory is particularly useful for describing collective
excitations on top of an unperturbed many-body background. In this work, we
consider harmonically trapped spinor BECs of which the initial state is
prepared in the SMA form, with the spin degrees of freedom in
far-from-equilibrium states. We develop a coupled-channel framework based on
the Bogoliubov theory to capture spin-spatial resonances. One challenge in
using Bogoliubov eigenmodes, particularly those with particle–hole
correlations, is that the particle number is not conserved. This is the case
because standard Bogoliubov theory is non-Hermitian and breaks the $U(1)$
symmetry~\cite{cd_lam}. These features make it challenging to describe the
long-time dynamics of systems with such a basis. Several approaches to overcome
this challenge have been discussed in the literature, such as consistently
including the back-action of the excitations on the background~\cite{48sm-thzm,
PhysRevA.106.053319, PhysRevD.72.105005} or introducing an additional mode that
corresponds to phase fluctuations of the background~\cite{PhysRevLett.77.3489}.
A modified version of standard Bogoliubov theory, which preserves the $U(1)$
symmetry, has also been developed in the literature~\cite{cd_lam,
PhysRevA.56.1414}. The framework developed in our work builds on the
$U(1)$-symmetry-preserving Bogoliubov theory of Refs.~\cite{cd_lam,
PhysRevA.56.1414}. While our paper focuses on the mean-field regime, extensions
to the beyond-mean-field regime are expected to follow an analogous derivation.

The remainder of the paper is organized as follows. Section II introduces the
system Hamiltonian, outlines the main logic behind the derivation of the
coupled-channel model developed in this work, and places our key results into
context. Section III reviews the Bogoliubov theory and presents the derivation
of the coupled-channel model and the resonance conditions. Finally, Section IV
presents conclusions and an outlook. Numerical details are relegated to
Appendix~\ref{app:numerics}.

\section{System overview and summary of key results}

\subsection{Mean-field theory for spin-1 BECs}
\label{sec:hamiltonian}

The mean-field energy functional for spin-1 BECs described by the spinor wave
function $\vpsi(\vr, t)
= \left[ \psi_{1}(\vr, t), \psi_0(\vr, t), \psi_{-1}(\vr, t) \right]^\trans$ is
\begin{equation}
\label{eq:mean_field_H}
H = \Hscal + \Hs,
\end{equation}
where the spin-independent energy functional $\Hscal$ and the spin-dependent
energy functional $\Hs$ are
\begin{align}\label{eq:H_sc}
\Hscal ={}& \int d\vr \vpsi^\dagger(\vr, t) \times
\nonumber \\
&\left[ - \frac{\hbar^2}{2 m} \nabla^2 + V(\vr)
+ \frac{g_n}{2} \abs*{\vpsi(\vr, t)}^2 \right] \vpsi(\vr, t)
\end{align}
and
\begin{multline}
\label{eq:H_S}
\Hs = \int d\vr \left\{ q \vpsi^\dagger(\vr, t) \vF_z^2 \vpsi(\vr, t)
\right.
\\
\left. + \frac{g_s}{2} \sum_{j=x,y,z} \left[ \vpsi^\dagger(\vr, t)
\vF_j \vpsi(\vr, t) \right]^2 \right\},
\end{multline}
respectively. Here, $V(\vr)$ denotes an external single-particle potential, the
spin-1 Pauli matrices $\vF_j$ take the forms
\begin{equation}
\begin{gathered}
\vF_x = \frac{1}{\sqrt{2}} \begin{pmatrix}
0 & 1 & 0 \\
1 & 0 & 1 \\
0 & 1 & 0
\end{pmatrix}
,\quad
\vF_y = \frac{1}{\sqrt{2}} \begin{pmatrix}
0 & -\ic & 0 \\
\ic & 0 & -\ic \\
0 & \ic & 0
\end{pmatrix}
, \\
\quad
\vF_z = \begin{pmatrix}
1 & 0 & 0 \\
0 & 0 & 0 \\
0 & 0 & -1
\end{pmatrix},
\end{gathered}
\end{equation}
and $m$, $\hbar$, and $q$ denote the atomic mass, the reduced Planck constant,
and the quadratic Zeeman shift, respectively. Using the normalization
\begin{equation}
\label{eq:mf_norm}
\int d\vr \vpsi^\dagger(\vr, t) \vpsi(\vr, t) = 1,
\end{equation}
the nonlinear coupling constants $g_n$ and $g_s$ for a 3D spinor BEC are
related to the spin-independent scattering length $a_n$, the spin-dependent
scattering length $a_s$, and the total particle number $N$ via
\begin{equation}
\label{eq:gns}
g_{n/s} = \frac{4 \pi \hbar^2 a_{n/s} (N-1)}{m}.
\end{equation}
For typical spinor BECs, the value of $g_s$ is much smaller than that of $g_n$
(e.g., $g_s /g_n \approx 1/28.10$ for $^{23}$Na~\cite{PhysRevA.83.042704}),
which leads to a scale separation between the dynamics of the spin degrees of
freedom and those of the spatial components. This scale separation underlies
the widely-used SMA treatment of spinor BECs~\cite{Ho_PhysRevLett1998,
Ho_PhysRevA2000, You_PhysRevA2005}.

At the mean-field level, the equation of motion for $\psi_j(\vr, t)$ is
determined by varying the mean-field energy functional $H$ [see
Eq.~\eqref{eq:mean_field_H}] with respect to $\psi^*_j(\vr, t)$, yielding the
spinor Gross-Pitaevskii equation (SGPE)~\cite{KAWAGUCHI2012253}
\begin{equation}\label{eq:sgpe}
\ic \hbar \frac{\partial \vpsi(\vr, t)}{\partial t}
= \cHsgp(\vr, t) \vpsi(\vr, t),
\end{equation}
with
\begin{equation}\label{eq:H_sgp}
\cHsgp(\vr, t) = \cH_\text{scalar}(\vr, t) + \cHs(\vr, t),
\end{equation}
\begin{equation}\label{eq:h_scalar}
\cHscal(\vr, t) = -\frac{\hbar^2\nabla^2}{2m} + V(\vr)
+ g_n \abs*{\vpsi(\vr, t)}^2,
\end{equation}
and
\begin{equation}
\label{eq:h_s}
\cHs(\vr, t)= q \vF_z^2 + g_s \sum_{j=x,y,z}
\left[\vpsi^\dagger(\vr, t) \vF_j \vpsi(\vr, t)\right] \vF_j.
\end{equation}
The SMA assumes that a common spatial wavefunction $\phi_0(\vr)$ is shared
by all spin components, i.e.,
\begin{equation}\label{eq:sma_ansatz}
\vpsi(\vr, t) \approx \phi_0(\vr) \vs_0(t),
\end{equation}
where $\vs_0(t)$ denotes the spin wavefunction and $\phi_0(\vr)$ is normalized
to $1$:
\begin{equation}
\int d\vr |\phi_0(\vr)|^2 =1.
\end{equation}
The spatial wavefunction $\phi_0(\vr)$ can be determined by minimizing the
scalar energy functional $\Hscal$.

We use $\cHscal^\text{SMA}(\vr)$ and $\cHs^\text{SMA}(\vr, t)$ to denote the
mean-field Hamiltonians $\cHscal(\vr, t)$ and $\cHs(\vr, t)$ under the SMA:
$\cHscal^\text{SMA}(\vr)$ and $\cHs^\text{SMA}(\vr, t)$ are obtained by
replacing $\vpsi(\vr, t)$ in Eqs.~\eqref{eq:h_scalar} and~\eqref{eq:h_s} with
$\phi_0(\vr)\vs_0(t)$. After this replacement, the only dynamical variable left
is the spin vector $\vs_0(t)$, for which the SMA equation of motion reads
\begin{equation}
\label{eq:spin_sma}
\ic \hbar \frac{\partial \vs_0(t)}{\partial t} = [\vhs(t) + \mun] \vs_0(t),
\end{equation}
where
\begin{equation}
\label{eq:spin_sma_h}
\vhs(t) = q \vF_z^2 + c_s\sum_{j=x,y,z}
\left[ \vs_0^\dagger(t) \vF_j \vs_0(t) \right] \vF_j,
\end{equation}
\begin{equation}
\mun = \int d\vr \phi_0^*(\vr) \cH^\text{SMA}_\text{scalar}(\vr) \phi_0(\vr),
\end{equation}
and
\begin{equation}
c_s = g_s \int d\vr |\phi_{0}(\vr)|^4.
\end{equation}

\subsection{Dynamics near resonances}
\label{sec:gp}

For many parameter combinations, the dynamics of an initial state of the form
given in Eq.~\eqref{eq:sma_ansatz} can, to a good approximation, be described
within the SMA. Under certain conditions, however, this approximation can break
down dynamically. For instance, as shown in Refs.~\cite{qth, qexp}, significant
resonantly enhanced spatial dynamics can arise at certain resonant values of
the quadratic Zeeman shift $q$, thereby breaking the SMA assumption.

In this section, we aim, through SGPE simulations, to identify the parameter
regimes in which the spinor BEC exhibits substantial spin-spatial resonant
dynamics. Throughout, the initial state is prepared to be of the SMA form given
in Eq.~\eqref{eq:sma_ansatz}, with $\phi_0(\vr)$ being the ground state of
$\cHscal(\vr, t)$. The system starts in a non-equilibrium spin state and
subsequently undergoes spin dynamics. We probe the spatial dynamics by
examining the density of a single hyperfine component as well as the total
density of the condensate. For illustrative purposes, we perform SGPE
simulations for 1D systems. The SGPE simulations reveal several resonances; the
underlying nature of these resonances is described in
Sec.~\ref{sec:bog_theory}.

\begin{figure*}[htbp]
\centering
\includegraphics[width=\textwidth]{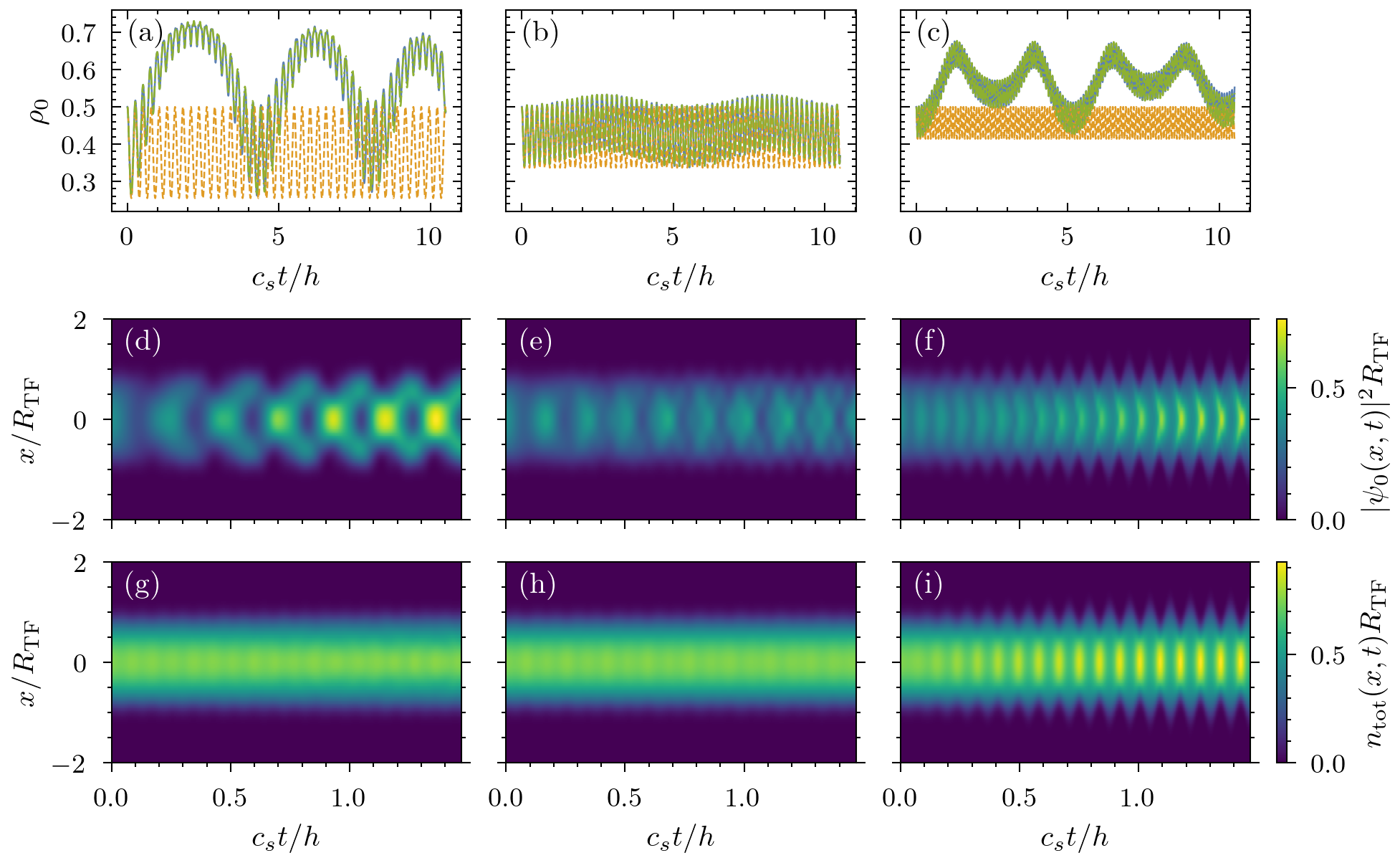}
\caption{
Near-resonant spin and spatial dynamics for a spin-1 BEC in a 1D harmonic trap
at quadratic Zeeman shifts $q = h \times 60$~Hz (first column),
$q = h \times 90$~Hz (middle column), and $q \approx h \times 171$~Hz (third
column). The first, second, and third rows show the time traces of $\rho_0(t)$,
the density $|\psi_0(x,t)|^2$ of the $m_F=0$ component, and the total density
$n_\text{tot}(x, t)=\vpsi^\dagger(x, t)\vpsi(x,t)$, respectively. The blue
solid lines, orange dashed lines, and green dash-dotted lines in panels (a)-(c)
correspond to the results of SGPE simulations, the SMA model, and the
$\vl$-model developed in this work with $19$ states, respectively. The SGPE and
$\vl$-model show near-identical dynamics, while the SMA fails to capture the
resonance-driven dynamics. The second and third rows show SGPE results. The
values of the angular trap frequency $\omega$, the Thomas-Fermi radius
$R_\text{TF}$, and the spin-dependent collision energy $c_s$ can be found in
the main text.
}
\label{fig:gp_ns}
\end{figure*}

As an example, Fig.~\ref{fig:gp_ns} shows SGPE simulation results for $^{23}$Na
atoms in a 1D harmonic trap with $V(x) = m \omega^2 x^2 / 2$ for the initial
spin state $\vs_0(0)=(\sqrt{0.25}, \sqrt{0.5}, \sqrt{0.25})^\trans$ for three
$q$ values, i.e., for $q = h \times 60$~Hz, $q = h \times 90$~Hz, and
$q \approx h \times 171$~Hz. To obtain the 1D-equivalents $g_{n/s,{\text{1D}}}$
of the coupling constants $g_n$ and $g_s$, we first choose an
experimentally-accessible scalar chemical potential $\mu_n$. We then use the 1D
expression for the density distribution of a scalar BEC given by the
Thomas-Fermi approximation and solve for $g_{n,{\text{1D}}}$, referred to from
here on as $g_n$ for notational convenience (there should not be any confusion
since all concrete examples in this work utilize a 1D geometry), under the
constraint that the density is normalized to unity. We obtain
\begin{equation}
g_n = \frac{4}{3} \sqrt{\frac{2 \mu_n^3}{m \omega^2}}.
\end{equation}
For $\omega = 2 \pi \times 196$~Hz and $\mun = 5.368 E_\text{ho}$, as used in
Fig.~\ref{fig:gp_ns}, this yields the scalar coupling constant
$g_n = 23.21 E_\text{ho} a_\text{ho}$ and the Thomas-Fermi radius
$R_\text{TF} = \sqrt{2 \mun/(m \omega^2)} = 3.276 a_\text{ho}$. The 1D
spin-dependent-coupling constant $g_s$ is subsequently set to
$g_s = g_n/28.10$, which yields the spin-dependent collision energy
$c_s = 0.1500 E_\text{ho}$. Here, $a_\text{ho}$ and $E_\text{ho}$ denote the
harmonic oscillator length and energy, respectively,
$a_\text{ho} = \sqrt{\hbar / (m \omega)}$ and $E_\text{ho} = \hbar \omega$. The
values of $\omega$, $c_s$, and $\mun$ lie in the typical parameter regime for
standard $^{23}$Na BECs prepared in the laboratory~\cite{Dan2023RevModPhys}.
The scale separation is satisfied since $\mun$ is much larger than $c_s$.

Figures~\ref{fig:gp_ns}(a)–\ref{fig:gp_ns}(c) show the time traces of the
population fraction $\rho_0(t)=\int dx |\psi_0(x, t)|^2$ of the $m_F=0$
component. Significant differences between the SGPE results (blue solid lines,
which lie underneath the green dash-dotted line) and the SMA results (orange
dashed lines) indicate a breakdown of the assumption that the dynamics is
governed by a single spatial orbital. The SGPE results exhibit slow
oscillations modulated by fast oscillations, indicating the presence of
multiple time scales. The fast oscillation with a period of roughly $0.25h/c_s$
is consistent with the SMA spin dynamics, while the slow oscillation, with a
much longer period of approximately $5.0h/c_s$, is determined by the coupling
between the background state, which is proportional to $\phi_0(x)$, and the
modes that are associated with spatial excitations. The green dot-dashed lines
in Figs.~\ref{fig:gp_ns}(a)-\ref{fig:gp_ns}(c), which reproduce the SGPE
results essentially exactly, correspond to the results obtained from the
beyond-SMA model, i.e., the $\vl$-model, which is newly developed in our work.

Figures~\ref{fig:gp_ns}(d)–\ref{fig:gp_ns}(f) show the spatial densities
$|\psi_0(x,t)|^2$ of the $m_F=0$ component for the same three $q$ values as
used in Figs.~\ref{fig:gp_ns}(a)–\ref{fig:gp_ns}(c). For $q = h \times 60$~Hz
[Fig.~\ref{fig:gp_ns}(d)], $|\psi_0(x,t)|^2$ exhibits oscillations between a
wavefunction with two peaks near the edges and a wavefunction with a single
peak at the center. In contrast, for $q \approx h \times 171$~Hz
[Fig.~\ref{fig:gp_ns}(f)], the dynamics of $|\psi_0(x,t)|^2$ display breathing
in which the condensate component expands and shrinks with non-negligible
density at $x=0$ throughout the dynamics, clearly distinct from the
$q = h \times 60$~Hz case. For $q = h \times 90$~Hz [Fig.~\ref{fig:gp_ns}(e)],
the spatial dynamics of $|\psi_0(x,t)|^2$ is much weaker and exhibits a mixed
behavior intermediate between the cases shown in Figs.~\ref{fig:gp_ns}(d)
and~\ref{fig:gp_ns}(f).

The distinct oscillatory behavior of $\rho_0$ at frequencies that deviate from
the SMA oscillation frequency can be attributed to spin-spatial resonances. To
further highlight the distinct features near these resonances,
Figs.~\ref{fig:gp_ns}(g)–\ref{fig:gp_ns}(i) show the total density
$n_\text{tot}(x,t) = \vpsi^\dagger(x,t) \vpsi(x,t)$ for the same three values
of $q$. For the resonances at $q = h \times 60$~Hz and $q = h \times 90$~Hz,
$n_\text{tot}(x,t)$ is approximately time independent. In contrast, for
$q \approx h \times 171$~Hz, $n_\text{tot}(x,t)$ exhibits pronounced breathing
dynamics similar to those of $|\psi_0(x,t)|^2$. Therefore, the former two
resonances correspond primarily to excitations within the spin degrees of
freedom, associated with oscillating spin structures across the condensate,
whereas the latter corresponds to a scalar density excitation.

\begin{figure*}[htbp]
\centering
\includegraphics[width=\textwidth]{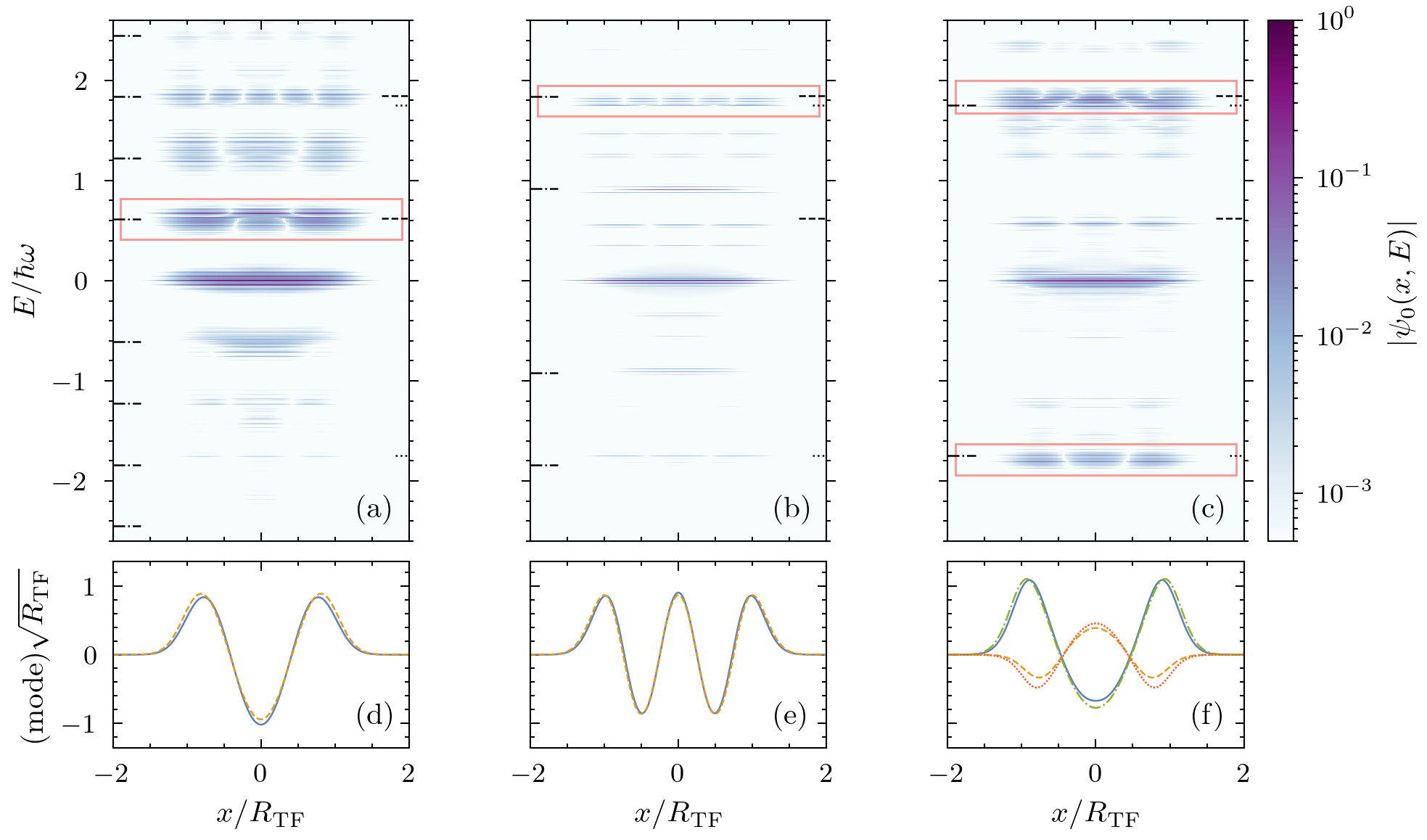}
\caption{
(a)-(c): The amplitudes $|\psi_0(x, E)|$ of the $m_F=0$ component of the spinor
wavefunction in the energy-space domain, normalized as a fraction of the peak
value [the legend on the right applies to all three panels; note that
$|\psi_0(x,E)|$ is---due to the chosen scaling---dimensionless]. The
parameters for the results in the left, middle, and right columns are the same
as those in Fig.~\ref{fig:gp_ns}. The dash-dotted lines on the left edges of
panels (a)–(c) indicate, in accordance with the resonance condition derived in
Sec.~\ref{sec:res_cond}, multiples of $2 q$. The dashed lines and dotted lines
on the right edges of panels (a)–(c) indicate the eigenenergies of the
Bogoliubov Hamiltonian derived from $\cH_\text{scalar}(\vr, t)$; the dashed
lines correspond to the spin-perpendicular excitations while the dotted lines
correspond to spin-parallel excitations (see the discussion in
Sec.~\ref{sec:bog_theory} for details). The red boxes in panels (a)-(c)
highlight the regions around the expected resonance positions. The second-most
occupied slices, which lie within the red boxes, have energies that coincide to
a good approximation with the estimated resonance condition. (d)-(f): Panels
(d) and (e) compare the real part of the wavefunctions $\psi_0(x,E)$ (blue
solid lines), selected from slices within the red boxes in (a) and (b), with
the real spatial part of the eigenmodes $u(x)$ and $v^*(x)$ of the Bogoliubov
Hamiltonian derived in Sec.~\ref{sec:bog_theory} for the spin-perpendicular
excitations (orange dashed lines). Panel (f) compares the real part of both the
maximally occupied positive-energy slice (blue solid line) and the
negative-energy slice (orange dashed line) with the corresponding Bogoliubov
spatial wavefunctions for the spin-parallel excitations [green dash-dotted line
for the positive-energy component $u(x)$ and red dotted line for the
negative-energy component $v^*(x)$]. The Bogoliubov wavefunctions and FFT
slices are normalized according to Eq.~(\ref{eq:norm_bdg}). The FFT slices show
good agreement with the Bogoliubov wavefunctions, although small discrepancies
are observed in panel (f) for the spatial part of the mode with particle–hole
correlations.
}
\label{fig:ffts}
\end{figure*}

To unambiguously identify the spatial parts of the modes involved in the
dynamics shown in Fig.~\ref{fig:gp_ns}, we compute the energy-space spinor
wavefunction $\psi_{m_F}(x,E)$ by performing a Fourier transform of
$\psi_{m_F}(x,t)$ over time,
\begin{align}
\psi_{m_F}(x, E)=
\frac{1}{\sqrt{2 \pi \hbar}} \int_{-\infty}^\infty
\psi_{m_F}(x, t)e^{-\frac{\ic E t}{\hbar}}dt .
\end{align}
Since the SGPE simulations provide $\psi_{m_F}(x,t)$ only within a finite time
window, we apply a Hamming window (see Appendix~\ref{app:numerics}) to mitigate
finite-window effects that arise from truncating the time integral at finite
limits. Figure~\ref{fig:ffts} shows $\psi_0(x,E)$ for the three near-resonant
$q$-values used in Fig.~\ref{fig:gp_ns}. It is evident that multiple spatial
wavefunctions with distinct energies $E$ contribute to the spectrum. We take
the energy slice with the dominant population, which corresponds to the ground
state of $\cH_\text{scalar}(\vr, t)$, as a reference and shift it to $E = 0$.
The red boxes in Figs.~\ref{fig:ffts}(a)-\ref{fig:ffts}(c) identify the
second-most occupied energy slices (the background state at $E = 0$ has the
highest occupation). For $q = h \times 60$~Hz, the second-most occupied energy
is $E \approx h \times 120$~Hz, which is approximately equal to $2 q$. In each
case, we refer to this second-most occupied energy as the resonant energy. For
$q = h \times 90$~Hz, the resonant energy is $E \approx h \times 360$~Hz, which
is approximately equal to $4 q$. For $q \approx h \times 171$~Hz, the resonant
energies are $E \approx \pm h \times 340$~Hz, which is approximately equal to
$\pm 2q$. A key difference is that, for the $q = h \times 60$~Hz and
$q = h \times 90$~Hz cases, resonant excitations appear only in the
positive-energy region, whereas for $q \approx h \times 171$~Hz resonant
excitations emerge in both the positive-energy region and the negative-energy
region. Correspondingly, the former two cases (with a resonant mode at positive
energy) correspond to resonances for which the modes lack particle–hole
correlations, while the latter case (positive and negative energy modes)
corresponds to resonances for which particle–hole correlations play a role.
Figures~\ref{fig:ffts}(d)-\ref{fig:ffts}(f) compare the resonant spatial
wavefunctions obtained by Fourier-transforming the SGPE results with the
spatial part of the eigenmodes of the Bogoliubov Hamiltonian derived in
Sec.~\ref{sec:bog_modes}. The agreement is very good. The observation that only
a few resonant spatial modes are involved, together with the very good
agreement between the FFT wavefunctions and the Bogoliubov eigenmodes,
indicates that Bogoliubov eigenmodes provide an effective basis for a
coupled-channel description of the resonance dynamics. These observations form
the backbone of the theory framework developed in this paper.

\subsection{Framing of our results}
\label{sec:summary_results}

This paper develops a coupled-channel framework based on a
particle-number-preserving Bogoliubov theory to describe the dynamics of spinor
BECs. Focusing on harmonically trapped systems, where well-defined spin-spatial
resonances exist, our framework is designed to accurately capture those
spin-spatial resonances, which dynamically drive the system out of the SMA
regime. Our Bogoliubov theory is based on the spin-independent part of the
spinor BEC Hamiltonian and does not incorporate any parameters associated with
the characteristic spin energy scales (i.e., the Bogoliubov modes are
independent of $c_s$ and $q$). Instead, $c_s$ and $q$ are accounted for through
the coupled-channel treatment of the spin dynamics and the expansion
coefficients of the Bogoliubov eigenmodes. Since the Bogoliubov eigenmodes
carry the spatial structure of the spinor, the resulting framework naturally
bridges the regime in which well-defined resonances exist and the regime in
which complex spatial structures, such as spin domains, emerge.

Aligned with our goal of describing the dynamical breakdown of the SMA, our
Bogoliubov theory is formulated around a vacuum state expressed as the direct
product of a spatial wavefunction and a time-dependent spinor. For the
spin-independent part of the spinor BEC Hamiltonian, the Bogoliubov spectrum
and the corresponding spatial components of the Bogoliubov eigenmodes are
independent of the background spin state and therefore time independent. The
eigenmodes subsequently serve as the channel functions in our coupled-channel
framework. We find that the Bogoliubov spectrum can be divided into two
uncoupled sectors: one with spin parallel to the background and the other with
spin perpendicular to the background. The former corresponds to excitations
with particle–hole correlations, while the latter corresponds to excitations
without such correlations. These findings uncover the nature of the two types
of resonances observed in Figs.~\ref{fig:gp_ns}(a) and \ref{fig:gp_ns}(c).

\begin{figure}[htbp]
\centering
\includegraphics[width=\linewidth]{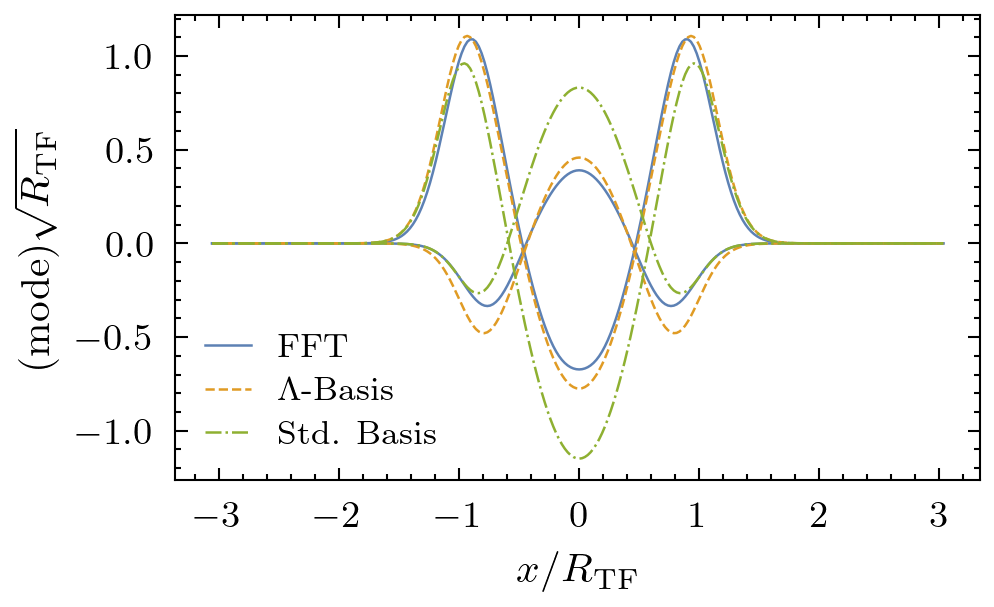}
\caption{
Comparison of the spatial part of the resonant mode obtained from the Fourier
transform of the SGPE results with the predictions of different Bogoliubov
theories for the parameters of Fig.~\ref{fig:ffts}(f). The blue solid lines
correspond to the real spatial parts of the maximally occupied positive-energy
and negative-energy excitations obtained by Fourier transforming the SGPE
results. The orange dashed lines correspond to the spatial part of the
particle-component of the positive-energy and negative-energy eigenmodes of the
$U(1)$-symmetry–conserving Bogoliubov theory developed in this work. The
additional green dash-dotted lines correspond to the spatial parts of the
positive-energy and negative-energy particle-components of the eigenmodes
obtained from the standard Bogoliubov theory that breaks the $U(1)$ symmetry.
The $U(1)$-symmetry–conserving theory provides a significantly better
description of the resonant modes than the standard Bogoliubov theory.
}
\label{fig:fft_mode}
\end{figure}

To overcome the non-hermiticity issue in the standard Bogoliubov theory that
leads to non-conserved normalization of the wavefunction and makes the
coupled-channel treatment diverge in the long-time limit, this work develops a
$U(1)$-symmetry-conserving Bogoliubov theory. The central idea is to restrict
the perturbations to the subspace orthogonal to the vacuum state. For modes
without particle–hole correlations, the spatial wavefunction of the standard
Bogoliubov mode is ``automatically'' orthogonal to the background state; in
this case, both Bogoliubov theories predict the same spatial parts of the
modes. For modes with particle–hole correlations, however, the eigenmodes of
our modified Bogoliubov theory differ from those of the standard Bogoliubov
theory. Figure~\ref{fig:fft_mode} compares the spatial part of the
particle-hole-correlated resonant mode, obtained from the FFT of the SGPE
results [also shown in Fig.~\ref{fig:ffts}(f)], with the spatial part of the
corresponding modes obtained from our modified Bogoliubov theory and the
standard Bogoliubov linearization of $\cHscal$ for the resonance near
$q = h \times 171$~Hz. It can be seen clearly that the modified Bogoliubov
theory (orange dashed lines) agrees with the FFT results (blue solid lines)
significantly better than the standard Bogoliubov theory (green dash-dotted
lines).

The extension developed in this work is related to the existing
$U(1)$-symmetry–preserving Bogoliubov theory developed in Refs.~\cite{cd_lam,
PhysRevA.56.1414}. In particular, both Bogoliubov theories contain a projection
operator that constrains the perturbations to the subspace orthogonal to the
vacuum state. Making an analogy to the approach developed in
Ref.~\cite{cd_lam}, their projection operator for the spinor system reads
\begin{equation}
\label{eq:projection_cd}
\vec{\cQ}'(\vr, \vr', t)
= \delta(\vr, \vr') - \vpsi(\vr, t) \vpsi^\dagger(\vr', t),
\end{equation}
where $\vpsi(\vr, t)$ describes the mean-field dynamics of the full system and
thus requires the full SGPE solution. In our system, in which the vacuum state
is characterized by a nearly frozen spatial profile and highly nonequilibrium
spin degrees of freedom, the Bogoliubov theory treats the projections onto the
spatial and spin subspaces separately. Specifically, we introduce an \emph{a
priori} spatial projection operator
\begin{equation}
\label{eq:projection_Q}
\cQ(\vr, \vr') = \delta(\vr, \vr') - \phi_0(\vr)\phi_0^*(\vr'),
\end{equation}
and a time-dependent spin projection operator
\begin{equation}
\cQs(t) = \Id - \vs_0(t) \vs_0^\dagger(t),
\end{equation}
which is determined on the fly within the coupled-channel framework. Here $\Id$
is the $3 \times 3$ identity matrix. Our approach of dividing the Hilbert space
and treating the spatial and spin degrees of freedom separately provides deep
insights into the physics near resonances. By analyzing the competition between
different energy scales and comparing with the full time evolution, our
treatment enables us to see the contributions of the terms neglected in
deriving the resonance conditions.

Section~\ref{sec:bog_theory} outlines our Bogoliubov theory and the resulting
``$\vl$-model'', namely, a coupled-channel model that provides a
$U(1)$-symmetric particle-conserving description of the dynamics near
resonances without the need of solving the full SGPE. Even though the exact
equations of motion for the mode amplitudes contain a multitude of terms due to
the nonlinearities of the system, the model is highly versatile in that it can
be truncated at various orders to examine the significance of different classes
of terms. For example, despite being motivated by Bogoliubov theory, the
$\vl$-model includes terms that go beyond standard Bogoliubov theory due to the
Bogoliubov modes serving as a basis. In Section~\ref{sec:res_cond}, we show
that a truncation of the $\vl$-model to an order comparable to what is
typically done in standard Bogoliubov theory, while retaining the coupling
between the background spin and the spatial excitation modes, explains the
resonance conditions of the SGPE. The ability to systematically vary the
truncation order of both the modes included and the terms retained in the
equations of motion provides a powerful tool for understanding the nature of
the excitations in the system and the significance of beyond-Bogoliubov effects
in the long-time dynamics.

\section{Bogoliubov theory and the \texorpdfstring{$\vl$}{Lambda}-model}
\label{sec:bog_theory}

\subsection{General formalism}
\label{sec:general_formalism}

This section seeks to elucidate the origin of the modes discussed in
Sec.~\ref{sec:gp} for a spinor Bose–Einstein condensate using a Bogoliubov
theory-based framework. To introduce perturbations about the background spatial
wavefunction $\phi_0(\vr)$, we write the mean-field spinor wavefunction as
\begin{equation}
\label{eq:psi_expansion}
\vpsi(\vr, t) = a_0(t) \phi_0(\vr) \vs_0(t) + \dvpsi(\vr, t),
\end{equation}
where the complex coefficient $a_0(t)$ describes the time evolution of the
population and phase of atoms occupying the spatial wavefunction $\phi_0(\vr)$,
$\vs_0(t)$ describes the internal spin dynamics, and $\dvpsi(\vr, t)$ accounts
for the dynamics of the perturbations. Importantly, the background spin
$\vs_0(t)$ is not, in general, taken to be the SMA solution, i.e., it is not
predetermined, but instead must be determined alongside the perturbation
$\dvpsi(\vr, t)$. If $\vs_0(t)$ was taken to be the SMA solution alongside a
predetermined $a_0(t)$, then Eq.~\eqref{eq:psi_expansion} would serve as the
starting point for developing what we are referring to as standard Bogoliubov
theory throughout this paper. We impose that $\dvpsi(\vr, t)$ is spatially
orthogonal to $\phi_0(\vr)$ for each spin component, i.e.,
\begin{equation}\label{eq:orthogonal_condition}
\int d\vr\, \phi_0^*(\vr) \dvpsi(\vr, t) = \vec{0}.
\end{equation}
With this choice, $\vs_0(t)$ fully captures the internal spin dynamics
associated with the spatial wavefunction $\phi_0(\vr)$, i.e., there is no
ambiguity in distinguishing the internal spin dynamics of the background state
from that of the excited spatial modes.

In what follows, we introduce separate projection operators for the spatial and
spin degrees of freedom. We define the spatial projection operator
$\cQ(\vr, \vr')$ via Eq.~\eqref{eq:projection_Q}. Using $\cQ(\vr, \vr')$, the
perturbed and background components of the spinor wavefunction can be written
as
\begin{equation}
\label{eq:perturbation_Q}
\dvpsi(\vr, t) = \int d\vr'\, \cQ(\vr, \vr') \vpsi(\vr', t)
\end{equation}
and
\begin{equation}
\label{eq:background_Q}
a_0(t) \phi_0(\vr) \vs_0(t) = \int d\vr'\, \left[\delta(\vr, \vr')
- \cQ(\vr, \vr')\right] \vpsi(\vr',t),
\end{equation}
respectively. This shows that the operator $\cQ(\vr, \vr')$ partitions the
spatial Hilbert space into the subspace spanned by $\phi_0(\vr)$ and the
subspace spanned by spatial functions orthogonal to $\phi_0(\vr)$.

We normalize $\vs_0(t)$ such that
\begin{equation}
\label{eq:normalization_s0}
\vs_0^\dagger(t) \vs_0(t)=1.
\end{equation}
There is a gauge freedom in choosing the overall phase of $\vs_0(t)$. It is
straightforward to verify that Eq.~\eqref{eq:normalization_s0} implies
$\dot{\vs}_0^\dagger(t) \vs_0(t) + \vs_0^\dagger(t) \dot{\vs}_0(t) = 0$, where
the dot above the background spin $\vs_0(t)$ denotes the time derivative. This
implies that $\dot{\vs}_0^\dagger(t) \vs_0(t)$ is purely imaginary. We
therefore impose the gauge condition
\begin{equation}
\label{eq:gauge_s0}
\dot{\vs}_0^\dagger(t) \vs_0(t) = 0.
\end{equation}
This choice implies that $a_0(t)$ describes the overall phase and the
population fraction of the background state $a_0(t) \phi_0(\vr) \vs_0(t)$,
while $\vs_0(t)$ describes the internal spin structure within this background
state.

The dynamics of $\dvpsi(\vr, t)$ are closely tied to the Bogoliubov spectrum of
perturbations about the background state $a_0(t) \phi_0(\vr) \vs_0(t)$. A major
challenge in the derivation is that this background state depends non-trivially
on time, which reduces the direct usefulness of the Bogoliubov spectrum for our
purposes. In a scalar system with a well-defined stationary background state,
the only time dependence of the background arises from the phase evolution that
is governed by the \emph{predetermined} chemical potential, which can be gauged
away. In contrast, in our treatment, as discussed above, neither the time
evolution of $a_0(t)$ nor that of $\vs_0(t)$ are known \emph{a priori};
instead, they are part of the solution to be determined. This makes it
impossible to fully rotate away the time dependence of the Bogoliubov
Hamiltonian.

To simplify the theory and establish a connection to Bogoliubov theory, we
leverage the characteristic time scales of $a_0(t)$ and $\vs_0(t)$. Since the
phase evolution of the background state is fully captured by $a_0(t)$, it is
natural to expect that $a_0(t)$ evolves on timescales that are set by the
scalar Hamiltonian, such as the timescale associated with the chemical
potential or the spin-independent mean-field energy. These timescales are much
shorter than the characteristic timescales of $\vs_0(t)$, which are set by the
spin-dependent Hamiltonian. These considerations suggest that the dynamics of
$\vs_0(t)$ are ``more adiabatic" than that of $a_0(t)$. Motivated by this
observation, we define the perturbation $\vl(\vr, t)$ as
\begin{equation}
\label{eq:lambda}
\vl(\vr, t) = a_0^*(t) \dvpsi(\vr, t).
\end{equation}
In terms of $\vl(\vr, t)$, the expansion of $\vpsi(\vr, t)$ becomes
\begin{equation}
\label{eq:psi_expand_lambda}
\vpsi(\vr, t) = a_0(t) \left[ \phi_0(\vr) \vs_0(t)
+ \frac{1}{N_0(t)} \vl(\vr, t) \right],
\end{equation}
where $N_0(t)=|a_0(t)|^2$ is the population of the background state. The key
benefit of working with $\vl(\vr, t)$ as opposed to $\dvpsi(\vr, t)$ is that
the redefinition of the perturbation [Eq.~\eqref{eq:psi_expand_lambda}]
enables, as will be shown below, the time-dependent phase of $a_0(t)$ be
removed from the equations of motion. Equation~\eqref{eq:psi_expand_lambda}
also clearly separates the phase dynamics and spin dynamics of the background
state, with the phase dynamics captured by $a_0(t)$ being much faster than the
spin dynamics captured by $\vs_0(t)$. The perturbation $\vl(\vr, t)$ then is
only sensitive to the slower time scales that are associated with the
population and spin structure of the background state. Our definition of
$\vl(\vr, t)$ bears close similarity to the perturbations introduced in
Refs.~\cite{cd_lam, PhysRevA.56.1414} and, to leading order, the Bogoliubov
Hamiltonian for $\vl(\vr, t)$ has an analogous form as the Bogoliubov
Hamiltonian obtained in particle–number–conserving Bogoliubov theory.

With these definitions in place, the main objective is to derive equations of
motion for $\vl(\vr, t)$, $a_0(t)$, and $\vs_0(t)$. One approach is to treat
$a_0(t)$, $\vs_0(t)$, and $\dvpsi(\vr, t)$ as independent dynamical variables
and to express the full Hamiltonian $H$ in terms of these dynamical variables.
From this representation, the equations of motion for $a_0(t)$, $\vs_0(t)$, and
$\dvpsi(\vr, t)$ can be derived. The equation of motion for $\vl(\vr, t)$
subsequently follows from Eq.~\eqref{eq:lambda} together with the identity
\begin{equation}
\frac{\partial \vl(\vr, t)}{\partial t} =
\frac{1}{a_0^*(t)} \frac{\partial a^*_0(t)}{\partial t} \vl(\vr, t)
+ a_0^{*}(t) \frac{\partial \dvpsi(\vr, t)}{\partial t}.
\end{equation}
For the gauge convention defined in Eq.~\eqref{eq:gauge_s0}, we find that the
equations of motion for $a_0(t)$, $\vs_0(t)$, and $\vl(\vr, t)$ are
\begin{equation}
\begin{aligned}[t]
\label{eq:eom_a0}
\ic \hbar \frac{\partial a_0(t)}{\partial t}
=& a_0(t) \vs_0^\dagger(t)
\int d\vr\phi^*_0(\vr)\cHsgp(\vr, t) \times \\
& \left[\phi_0(\vr)\vs_0(t) + \frac{\vl(\vr,t)}{N_0(t)}\right],
\end{aligned}
\end{equation}
\begin{equation}
\begin{aligned}[t]
\label{eq:eom_s0}
\ic \hbar \frac{\partial \vs_0(t)}{\partial t}
= &\cQs(t) \int d\vr \phi^*_0(\vr)\cHsgp(\vr, t) \times \\
& \left[\phi_0(\vr)\vs_0(t) + \frac{\vl(\vr,t)}{N_0(t)}\right],
\end{aligned}
\end{equation}
and
\begin{multline}
\label{eq:eom_lambda}
\ic \hbar \frac{\partial \vl(\vr, t)}{\partial t} =
\\
\begin{aligned}[b]
&N_0(t) \int d\vr' \cQ(\vr, \vr') \cHsgp(\vr', t) \phi_0(\vr') \vs_0(t)
\\
+& \int d\vr' \cQ(\vr, \vr') \cHsgp(\vr', t)
\vl(\vr',t)
\\
-& \left[ \int d\vr' \vs_0^\dagger(t) \phi^*_0(\vr')
\cHsgp(\vr', t) \phi_0(\vr') \vs_0(t) \right]\vl(\vr, t)
\\
-& \left[ \int d\vr'\vl^\dagger(\vr',t)
\cHsgp(\vr', t) \phi_0(\vr')\vs_0(t) \right] \frac{\vl(\vr, t)}{N_0(t)}.
\end{aligned}
\end{multline}
In Eqs.~\eqref{eq:eom_a0}–\eqref{eq:eom_lambda}, it is implied that the
wavefunction $\vpsi(\vr, t)$ that is contained in $\cHsgp(\vr, t)$ [see
Eqs.~\eqref{eq:H_sgp}-\eqref{eq:h_s}] is given by
Eq.~\eqref{eq:psi_expand_lambda}.
Equations~\eqref{eq:eom_a0}–\eqref{eq:eom_lambda} form a closed set of coupled
differential equations for the three dynamical variables.

The first term on the right-hand side of Eq.~\eqref{eq:eom_lambda}, which is
independent of $\vl(\vr, t)$, acts as a seeding term for the growth of
$\vl(\vr, t)$. This term is nonzero provided $\phi_0(\vr) \vs_0(t)$ is, at
fixed time $t$, not an eigenstate of $\cHsgp(\vr, t)$ and the spatial integral
does not vanish accidentally. Since $\phi_0(\vr) \vs_0(t)$ is, for
far-from-equilibrium states, not an eigenstate of $\cHsgp(\vr, t)$, coupling to
excited modes is, in general, present if the spatial degrees of freedom, the
spin degrees of freedom, or both are out of equilibrium. One example in which
the spatial integral in the seeding term vanishes is the homogeneous system,
where $\cHsgp(\vr', t)$ is independent of $\vr'$. In this case,
$\cHsgp(\vr', t)\phi_0(\vr')$ is proportional to $\phi_0(\vr')$, which lies
entirely in the condensate subspace and is therefore projected out by the
spatial projector $\cQ(\vr, \vr')$.

\subsection{Discussion on \texorpdfstring{$U(1)$}{U(1)} symmetry conservation}
\label{sec:u1_symmetry}

By discretizing the perturbation $\vl(\vr, t)$ on a spatial grid, the coupled
differential equations can be solved numerically, yielding results that are
fully equivalent to those obtained by solving the SGPE. One advantage of the
formalism summarized in Eqs.~\eqref{eq:eom_a0}–\eqref{eq:eom_lambda} is that it
allows us to treat the background and the excitations separately, thereby
facilitating a clear conceptual understanding of the nature of spin-spatial
resonances and providing a route to engineer resonance conditions. Another
technical advantage is that it allows the perturbation $\vl(\vr, t)$ to be
expanded in a set of well-defined basis functions, namely the Bogoliubov modes,
with the potential to capture the main features of the dynamics using a
relatively small number of modes. In standard Bogoliubov theory, the
basis-expansion approach breaks the system's $U(1)$ symmetry and therefore
suffers from the inability to preserve wavefunction normalization during the
time evolution. As discussed below, the formalism developed here resolves this
issue.

First, it can be shown that Eq.~\eqref{eq:eom_s0} preserves both the
normalization of $\vs_0(t)$ and the gauge convention defined in
Eq.~\eqref{eq:gauge_s0}, provided that the spin projection operator $\cQs(t)$
is retained on the right-hand side of the equation. Second,
Eq.~\eqref{eq:eom_lambda} can be shown to preserve the orthogonality between
$\vl(\vr, t)$ and $\phi_0(\vr)$, provided the spatial projection operator
$\cQ(\vr, \vr')$ is retained in the first two terms on the right-hand side of
the equation and $\cHsgp(\vr, t)$ is Hermitian. Third, since $\cHsgp(\vr, t)$
depends on $N_0(t)$ rather than $a_0(t)$, the equations of motion for
$\vs_0(t)$ and $\vl(\vr, t)$ [Eqs.~\eqref{eq:eom_s0} and~\eqref{eq:eom_lambda}]
depend on $N_0(t)$ but not on the phase of $a_0(t)$, i.e., the phase of the
background state. This $U(1)$-symmetry–conserving property of the theory
implies that the normalization of the wavefunction $\vpsi(\vr, t)$ is conserved
even if $\vl(\vr, t)$ is expanded in terms of a finite number of modes (an
incomplete basis) as opposed to a complete basis.

Using conservation of the normalization of $\vpsi(\vr, t)$,
\begin{equation}
\label{eq:normalization}
N_0(t) + \frac{1}{N_0(t)} \int d\vr \vl^\dagger(\vr, t) \vl(\vr, t) = 1,
\end{equation}
$N_0(t)$ can be eliminated from Eqs.~\eqref{eq:eom_s0}
and~\eqref{eq:eom_lambda} by solving Eq.~(\ref{eq:normalization}), assuming
$N_0(t) > 1/2$, for $N_0(t)$,
\begin{equation}
\label{eq:N0}
N_0(t) = \frac{1}{2} + \frac{1}{2} \left[ 1
- 4\int d\vr \vl^\dagger(\vr, t) \vl(\vr, t) \right]^{1/2}.
\end{equation}
Therefore, the dynamics of $\vs_0(t)$ and $\vl(\vr, t)$ can be determined
without solving for $a_0(t)$. Given $\vl(\vr, t)$, $N_0(t)$ can be obtained
from Eq.~\eqref{eq:N0}. Since $|a_0(t)| = \sqrt{N_0(t)}$, it follows that
$a_0(t)$ is determined up to an arbitrary phase. Since this overall phase does
not enter into any physical observable, it is irrelevant for our purposes. It
follows that there is no need to solve for $a_0(t)$ order by order, consistent
with the approximations used for determining $\vs_0(t)$ and $\vl(\vr, t)$.
Instead, the $U(1)$ symmetry guarantees that determining $|a_0(t)|$ from the
normalization condition yields a solution that is automatically accurate to the
same order as $\vs_0(t)$ and $\vl(\vr, t)$. In the following analysis, we
therefore focus on the equations of motion for $\vs_0(t)$ and $\vl(\vr, t)$.

\subsection{Connection to ``standard'' Bogoliubov theory}
\label{sec:bog_modes}

To make the connection to standard Bogoliubov theory explicit, we Taylor expand
the right-hand side of Eq.~\eqref{eq:eom_s0} up to linear order in $\vl(\vr,
t)$. This yields
\begin{multline}
\label{eq:eom_s0_expansion}
\ic \hbar \frac{\partial \vs_0(t)}{\partial t}
\approx \left[ \vhs(t) - \mus(t) \right] \vs_0(t)
+ g_s \cQs(t) \times
\\
\Bigg[
\sum_{j=x,y,z} \vs_0^\dagger(t) \vF_j \vs_0(t) \vF_j
\int d\vr \phi_0^*(\vr) \abs{\phi_0(\vr)}^2 \vl(\vr, t)
\\
+ \sum_{j=x,y,z} \vF_j \vs_0(t) \vs_0^\dagger(t) \vF_j
\int d\vr \phi_0^*(\vr) \abs{\phi_0(\vr)}^2 \vl(\vr, t)
\\
+ \sum_{j=x,y,z} \vF_j \vs_0(t) \vs_0^\trans(t) \vF_j^\trans
\int d\vr \phi^*_0(\vr) \phi_0^2(\vr) \vl^*(\vr, t)
\Bigg],
\end{multline}
where the time-dependent spin chemical potential $\mus(t)$ is given by
\begin{equation}
\mus(t) = \vs_0^\dagger(t) \vhs(t) \vs_0(t).
\end{equation}
Doing the same for Eq.~\eqref{eq:eom_lambda}, we find
\begin{multline}
\label{eq:eom_vl_lgp}
\ic \hbar \frac{\partial}{\partial t}
\begin{bmatrix} \vl(\vr, t) \\ \vl^*(\vr, t) \end{bmatrix}
\approx
\\
\vSl(\vr, t) + \int d\vr' \cLgp(\vr, \vr', t)
\begin{bmatrix} \vl(\vr', t) \\ \vl^*(\vr', t)
\end{bmatrix},
\end{multline}
where the driving term $\vSl(\vr, t)$ is given by
\begin{equation}
\label{eq:driving_term}
\vSl(\vr, t)=
\left[ \cHs^\text{SMA}(\vr, t) - \vhs(t) \right]
\begin{bmatrix}
\phi_0(\vr)\vs_0(t) \\
- \phi^*_0(\vr)\vs^*_0(t)
\end{bmatrix}.
\end{equation}
The full Bogoliubov Hamiltonian $\cLgp(\vr, \vr',t)$ contains the scalar part
$\cLscal(\vr, \vr',t)$ and the spin part $\cLs(\vr, \vr',t)$,
i.e., $\cLgp(\vr, \vr',t)=\cLscal(\vr, \vr',t)+\cLs(\vr, \vr',t)$,
where
\begin{widetext}
\begin{align}
\label{eq:bogoliubov_scalar}
&\cLscal(\vr, \vr',t) = \\ \nonumber
&\begin{bmatrix}
\left[\cH_\text{scalar}^\text{SMA}(\vr) - \mun\right] \delta(\vr, \vr')
+ g_n \left(\cQ \left||\phi_0|^2\right| \cQ\right) \vs_0(t) \vs_0^\dagger(t)
& g_n \left(\cQ\left|\phi^2_0\right| \cQ^*\right) \vs_0(t)\vs^\trans_0(t)
\\
- g_n \left(\cQ^*\left| \phi^{*2}_0\right|\cQ\right)
\vs^*_0(t) \vs_0^\dagger(t)
& -\left[\cH_\text{scalar}^\text{SMA}(\vr) - \mun\right] \delta(\vr, \vr')
- g_n \left(\cQ^*\left||\phi_0|^2\right| \cQ^*\right)
\vs^*_0(t) \vs^\trans_0(t)
\end{bmatrix}
\end{align}
and
\begin{align}
\label{eq:bogoliubov_s}
\cLs(\vr, \vr',t) = &
\begin{bmatrix}
\left( \cQ \left| \cHs^\text{SMA}(t) \right| \cQ \right) - \mus(t)
& 0 \\
0 & - \left( \cQ^* \left| \cHs^\text{SMA}(t) \right| \cQ^* \right)
+ \mus(t)
\end{bmatrix}\delta(\vr, \vr')
\\ \nonumber
&+ g_s
\begin{bmatrix}
\left(\cQ\left| |\phi_0|^2\right| \cQ \right)
\sum_{j=x,y,z} \vF_j \vs_0(t) \vs_0^\dagger(t) \vF_j
&\left(\cQ\left| \phi_0^2\right| \cQ^*\right)
\sum_{j=x,y,z} \vF_j \vs_0(t) \vs^\trans_0(t) \vF_j^\trans
\\
- \left(\cQ^*\left| \phi_0^{*2}\right| \cQ\right)
\sum_{j=x,y,z} \vF^\trans_j \vs^*_0(t)\vs_0^\dagger(t) \vF_j
& - \left(\cQ^*\left| |\phi_0|^2\right| \cQ^* \right)
\sum_{j=x,y,z} \vF_j
\vs^*_0(t) \vs^\trans(t) \vF_j
\end{bmatrix}.
\end{align}
\end{widetext}
Here, expressions of the form $\left(\cdots\left|\cdots\right|\cdots\right)$
should be understood as the $(\vr, \vr')$ matrix element of the triple matrix
product; e.g., $(\cQ\left| |\phi_0|^2\right| \cQ) = \int d\vr'' \cQ(\vr,
\vr'')|\phi_0(\vr'')|^2\cQ(\vr'', \vr')$. Both $\cLscal(\vr, \vr',t)$
and $\cLs(\vr, \vr',t)$ are time dependent due to the time dependence of
$\vs_0(t)$.

To get rid of the time dependence of $\cLscal(\vr, \vr',t)$, we move
to the rotating frame of $\vs_0(t)$. Note that the Taylor-expanded equation of
motion for $\vs_0(t)$ [Eq.~\eqref{eq:eom_s0_expansion}] is governed by terms of
different orders in $\vl(\vr, t)$, with the leading- and sub-leading-order
terms proportional to $\vhs(t)-\mus(t)$ [this term is proportional to $g_s$ and
independent of $\vl(\vr, t)$] and $g_s \vl(\vr, t)$ [or $g_s \vl^*(\vr, t)$],
respectively. Since both $g_s$ and $\vl(\vr, t)$ are small, we neglect the $g_s
\vl(\vr, t)$ and $g_s \vl^*(\vr, t)$ terms (for our current purposes, we are
counting them as being effectively of quadratic order) and find approximately
\begin{equation}
\vs_0(t)=\vUsma(t) \vs_0(0),
\end{equation}
where
\begin{equation}\label{eq:U_SMA}
\vUsma(t) =
\exp\left( - \ic \int_0^t
\left[ \vhs(t') - \mus(t') \right] dt' / \hbar \right),
\end{equation}
i.e., the time evolution is, at this order, governed by the SMA solution to
$\vs_0(t)$. In the rotating frame of $\vs_0(t)$, we define the rotated
perturbation $\tvl(\vr, t)$ via
\begin{equation}
\tvl(\vr, t) = \left[\vUsma(t)\right]^{-1} \vl(\vr, t).
\end{equation}
The corresponding equation of motion up to linear order in $\tvl(\vr, t)$ is
\begin{multline}
\label{eq:eom_hl_lgp_rotated}
\ic \hbar \frac{\partial}{\partial t}
\begin{bmatrix} \tvl(\vr, t) \\ \tvl^*(\vr, t)
\end{bmatrix}
\approx
\\
\begin{aligned}[b]
\tvSl(\vr, t)
- \left[ \vhs(t) - \mus(t) \right]
&\begin{bmatrix} \tvl(\vr, t) \\ -\tvl^*(\vr, t) \end{bmatrix}
\\
+ \int d\vr' \tcLgp(\vr, \vr',t)
&\begin{bmatrix} \tvl(\vr', t) \\ \tvl^*(\vr', t) \end{bmatrix},
\end{aligned}
\end{multline}
where the rotated driving term is given by
\begin{equation}
\label{eq:driving_term_rotated}
\tvSl(\vr, t) = \begin{bmatrix}
\tilde{\vec{\mathscr{H}}}(\vr, t) \phi_0(\vr) \vs_0(0)
\\
- \tilde{\vec{\mathscr{H}}}^*(\vr, t) \phi_0^*(\vr) \vs_0^*(0)
\end{bmatrix}
\end{equation}
with
\begin{multline}
\tilde{\vec{\mathscr{H}}}(\vr, t)
= \left[\vUsma(t)\right]^{-1} \left[
\cHs^\text{SMA}(\vr, t) - \vhs(t) \right] \times
\\
\vUsma(t).
\end{multline}
The rotated full Bogoliubov Hamiltonian consists of the rotated scalar part and
the rotated spin part, i.e.,
$\tcLgp(\vr, \vr',t) = \tcLscal(\vr, \vr') + \tcLs(\vr, \vr',t)$. The rotated
scalar part $\tcLscal(\vr, \vr')$ has the same form as
Eq.~\eqref{eq:bogoliubov_scalar} with $\vs_0(0)$ representing the spin state at
$t=0$. The rotated spin part $\tcLs(\vr, \vr',t)$ is obtained using the same
strategy as the one we used to obtain Eq.~\eqref{eq:bogoliubov_s}. In addition,
$\mus(t)$ is replaced by $\vhs(t)$. A key observation is that
$\cHs^\text{SMA}(\vr, t ) - \vhs(t)$ is independent of single-particle terms
such as the quadratic Zeeman shift term. Because of this, the contribution to
the equation of motion of $\tvl(\vr, t)$ from the rotated spin Bogoliubov
Hamiltonian is of the order $g_s \tvl(\vr, t)$, which can be---consistent with
our earlier counting of orders---neglected. With these approximations, the
leading order contribution of the quadratic Zeeman shift $q$ is incorporated
into the $\tilde{\vSl}(\vr, t)$ term, and the structure of $\tvl(\vr, t)$ is
determined by the time-independent Bogoliubov Hamiltonian $\tcLscal(\vr, \vr')$
with the time-dependent driving term $\tvSl(\vr, t)$, i.e.,
\begin{multline}
\label{eq:eom_hl_lgp_rotated_approx}
\ic \hbar \frac{\partial}{\partial t}
\begin{bmatrix} \tvl(\vr, t) \\ \tvl^*(\vr, t) \end{bmatrix}
\approx
\\
\tvSl(\vr, t) + \int d\vr' \tcLscal(\vr, \vr')
\begin{bmatrix} \tvl(\vr', t) \\ \tvl^*(\vr', t) \end{bmatrix}.
\end{multline}

The eigenmodes of $\tcLscal(\vr, \vr')$ are obtained by solving
\begin{equation}
\label{eq:eigen_bogoliubov_rotated}
\int d\vr' \tcLscal(\vr, \vr')
\begin{bmatrix} \tvu_k(\vr') \\ \tvv_k(\vr')
\end{bmatrix}
= E_k \begin{bmatrix} \tvu_k(\vr) \\ \tvv_k(\vr)
\end{bmatrix},
\end{equation}
where $\tvu(\vr)$ and $\tvv(\vr)$ are spin-spatial wavefunctions that
represent, respectively, the Bogoliubov particle-component and Bogoliubov
hole-component. Note that $\tcLscal(\vr, \vr')$ depends on the initial
background spin state $\vs_0(0)$ through the spin operator
$\vs_0(0) \vs_0^\dagger(0)$; off-diagonal elements of $\tcLscal(\vr, \vr')$
vanish for a perturbation that is proportional to a spinor that is orthogonal
to $\vs_0(0)$. The Bogoliubov eigenmodes for spin-perpendicular excitations
therefore satisfy $\tvv_k(\vr)=0$. The presence of the spin operator enables a
decomposition of the Bogoliubov Hamiltonian into two subspaces, i.e.,
\begin{align}
\tcLscal(\vr, \vr') &= \tcLperp(\vr, \vr')
\begin{bmatrix}
\cQs(0) & 0\\
0 & \cQs^*(0)
\end{bmatrix}\\\nonumber
&+
\begin{bmatrix}
\vs_0(0) & 0\\
0 & \vs^*_0(0)
\end{bmatrix}
\tcLpar(\vr,\vr')
\begin{bmatrix}
\vs_0^\dagger(0) & 0\\
0 & \vs^\trans_0(0)
\end{bmatrix},
\end{align}
where the blocks $\tcLperp(\vr, \vr')$ and
$\tcLpar(\vr, \vr')$ for perpendicular and parallel spin states are,
respectively, given by
\begin{multline}
\label{eq:L_perp}
\tcLperp(\vr, \vr') =
\\
\begin{bmatrix}
\cH^\text{SMA}_\text{scalar}(\vr) - \mun & 0
\\
0 & - \left[ \cH^\text{SMA}_\text{scalar}(\vr) - \mun\right]
\end{bmatrix}\delta(\vr,\vr')
\end{multline}
and
\begin{widetext}
\begin{align}
\tcLpar(\vr, \vr') = \begin{bmatrix}
\left[\cH_\text{scalar}^\text{SMA}(\vr) - \mun\right] \delta(\vr, \vr')
+ g_n \left(\cQ \left||\phi_0|^2\right| \cQ\right)
& g_n \left(\cQ\left|\phi^2_0\right| \cQ^*\right)
\\
- g_n \left(\cQ^*\left| \phi^{*2}_0\right|\cQ\right)
& -\left[\cH_\text{scalar}^\text{SMA}(\vr) - \mun \right]\delta(\vr, \vr')
- g_n \left(\cQ^*\left||\phi_0|^2\right| \cQ^*\right)
\end{bmatrix}.
\end{align}
\end{widetext}
The blocks $\tcLperp(\vr, \vr')$ and
$\tcLpar(\vr, \vr')$ are diagonalized separately by
\begin{equation}\label{eq:eigenval_perp}
\int d\vr' \tcLperp(\vr,\vr')
\begin{bmatrix} \modeperp{k}(\vr') \\ 0 \end{bmatrix}
= E_{\perp k} \begin{bmatrix} \modeperp{k}(\vr) \\ 0 \end{bmatrix}
\end{equation}
and
\begin{equation}\label{eq:eigenval_para}
\int d\vr' \tcLpar(\vr, \vr')
\begin{bmatrix} \modeppar{k}(\vr') \\ \modehpar{k}(\vr') \end{bmatrix}
= E_{\parallel k}
\begin{bmatrix} \modeppar{k}(\vr) \\ \modehpar{k}(\vr) \end{bmatrix},
\end{equation}
where $\modeperp{k > 0}(\vr)$, $\modeppar{k > 0}(\vr)$, and
$\modehpar{k > 0}(\vr)$ are
scalar functions. Here we choose the normalization
\begin{equation}\label{eq:normalization_modeperp}
\int d\vr\, \abs*{\modeperp{k > 0}(\vr)}^2 = 1
\end{equation}
and
\begin{equation}
\label{eq:norm_bdg}
\int d\vr \left[|\modeppar{k}(\vr)|^2 - |\modehpar{k}(\vr)|^2\right] = 1.
\end{equation}

The decomposition of $\tcLscal(\vr, \vr')$ into two subspaces motivates us to
decompose $[\tvl(\vr, t), \tvl^*(\vr, t)]^\trans$ via
\begin{equation}
\label{eq:lambda_spin_decomp}
\begin{bmatrix}
\tvl(\vr, t) \\
\tvl^*(\vr, t)
\end{bmatrix}=
\begin{bmatrix}
\tvlp(\vr, t) \\
\tvlp^*(\vr, t)
\end{bmatrix}+
\begin{bmatrix}
\tvls(\vr, t) \\
\tvls^*(\vr, t)
\end{bmatrix}
,
\end{equation}
where the perpendicular and parallel excitations can be further expanded into
the two types of Bogoliubov eigenmodes via
\begin{equation}
\label{eq:bogoliubov_eigenmode_perp}
\begin{bmatrix} \tvlp(\vr, t) \\ \tvlp^*(\vr, t) \end{bmatrix}
= \sum_{k>0} \begin{bmatrix} \modeperp{k}(\vr) \tvs_k(t) \\ 0 \end{bmatrix}
+ \begin{bmatrix} 0 \\ \modeperp{k}^*(\vr) \tvs^*_k(t) \end{bmatrix}
\end{equation}
and
\begin{equation}
\begin{aligned}
\label{eq:bogoliubov_eigenmode_para}
\begin{bmatrix} \tvls(\vr, t) \\ \tvls^*(\vr, t) \end{bmatrix}
= \sum_{k>0} & \tilde{b}_{k}(t) \begin{bmatrix}
u_{\parallel k}(\vr) \vs_0(0)
\\
v_{\parallel k}(\vr) \vs^*_0(0)
\end{bmatrix}
\\
+
& \tilde{b}^*_k(t) \begin{bmatrix}
v^*_{\parallel k}(\vr) \vs_0(0)
\\
u^*_{\parallel k}(\vr) \vs^*_0(0)
\end{bmatrix}.
\end{aligned}
\end{equation}
The index $k$ in Eqs.~\eqref{eq:bogoliubov_eigenmode_perp}
and~\eqref{eq:bogoliubov_eigenmode_para} runs only over positive values because
contributions from the positive-energy and negative-energy eigenmodes are
included explicitly in each term of the sum.

In Eq.~\eqref{eq:bogoliubov_eigenmode_para}, $\tilde{b}_k(t)$ and
$\tilde{b}^*_k(t)$ are the complex amplitudes of the basis functions
$[\modeppar{k}(\vr) \vs_0(0), \modehpar{k}(\vr) \vs^*_0(0)]^\trans$ for the
spin-parallel subspace and the $\tvs_k(t)$ are the unnormalized time-dependent
complex spinor amplitudes of the basis functions $\modeperp{k > 0}(\vr)$ for
the spin-perpendicular subspace with the quantity
$\tvs_{k > 0}^\dagger(t) \tvs_{k > 0}(t)/ N_0(t)$ representing the occupation
fraction of the spin-perpendicular excited state labeled by $k$. The
time-dependent scalar amplitudes $\tilde{b}_{k>0}(t)$ are related to
$\tvl(\vr, t)$ via
\begin{align}
\tilde{b}_k(t) &= \int \vs_0^\dagger(0) \tvl(\vr, t) \modeppar{k}^*(\vr) d\vr
\\\nonumber
&- \int \tvl^\dagger(\vr, t) \vs_0(0) \modehpar{k}^*(\vr) d\vr, \qquad k > 0,
\end{align}
and the time-dependent spinor amplitudes $\tvs_{k > 0}(t)$ are related to
$\tvl(\vr, t)$ via
\begin{eqnarray}
\tvs_{k}(t) = \cQs(0) \int d\vr \modeperp{k}^*(\vr) \tvl(\vr, t), \qquad k > 0.
\end{eqnarray}
By construction, we have
\begin{equation}\label{eq:spin_orthogonality}
\tvs_{k > 0}^\dagger(t) \vs_0(0) = 0.
\end{equation}

Note that $\modeperp{k}(\vr)$ can be understood as an excitation in the
effective potential $V(\vr) + g_n|\phi_0(\vr)|^2-\mun$, consistent with the
effective potential picture introduced in Ref.~\cite{qth}. Since there exist
two spin projections for a spin-1 system that are orthogonal to $\vs_0(0)$,
there exist two degenerate modes with different spins that are associated with
each spatial wavefunction $\modeperp{k > 0}(\vr)$. With time evolution,
$\tvs_{k > 0}(t)$ may precess in the spin plane that is orthogonal to
$\tvs_0(0)$, hence the explicit time dependence. Since the excitations
$\tvls(\vr, t)$ are, in spin space, parallel to $\vs_0(0)$, the pair
$[\modeppar{k}(\vr), \modehpar{k}(\vr)]^\trans$ corresponds to excitations of a
scalar BEC obtained using $U(1)$-symmetry-preserving Bogoliubov
theory~\cite{cd_lam}. Because of the projection operators $\cQ(\vr, \vr')$ and
$\cQ^*(\vr, \vr')$ that appear in each element of $\tcLpar(\vr, \vr')$, the
spatial wavefunctions of the particle-hole coupled modes are orthogonal to the
background spatial wavefunction, i.e.,
\begin{align}
\int d\vr \phi_0^*(\vr) \modeppar{k > 0}(\vr) &= 0,
\\
\int d\vr \phi_0(\vr) \modehpar{k > 0}(\vr) &= 0.
\end{align}
This is in contrast to standard Bogoliubov theory~\cite{RevModPhys.73.307},
where such orthogonality does not, in general, exist. Since the scalar
wavefunctions $\phi_0(\vr)$ and $\modeperp{k}(\vr)$ are non-degenerate
eigenfunctions of the same scalar single-particle Hamiltonian, they must all be
orthogonal to each other as well, and thus
Eqs.~\eqref{eq:lambda_spin_decomp}-\eqref{eq:bogoliubov_eigenmode_para} satisfy
the orthogonality conditions set by Eq.~\eqref{eq:orthogonal_condition},
namely, $\int d\vr \phi_0^*(\vr) \vl(\vr, t) = 0$ for all excitations in the
system. The energy-space results shown in Fig.~\ref{fig:fft_mode} confirm that
these spatial amplitudes that obey the orthogonality condition just discussed
better match the numerics than those without it, suggesting that our extension
of the standard Bogoliubov theory better captures the properties of the
particle-hole-correlated excitations.

A complication arises in these modes in that the bases formed by them are not
orthogonal, and thus spatial overlaps between the modes of the \emph{different}
excitation types do not vanish:
\begin{align}
\label{eq:mode_overlap_p}
\int d\vr \modeperp{k > 0}^*(\vr) \modeppar{k' > 0}(\vr) &\neq 0,
\\
\label{eq:mode_overlap_h}
\qquad \int d\vr \modeperp{k > 0}(\vr) \modehpar{k' > 0}(\vr) &\neq 0.
\end{align}
However, since by definition the Bogoliubov expansion diagonalizes the dominant
Bogoliubov Hamiltonian $\tcLscal(\vr, \vr')$, the coupling between the states
of different bases only arises from the excitations of the low-energy spin
Bogoliubov Hamiltonian $\tcLs(\vr, \vr')$ as well as terms of higher order in
$\vl(\vr, t)$ that only become important at later times when the depletion of
the background state is appreciable.

\begin{figure}[b]
\centering
\includegraphics[width=\linewidth]{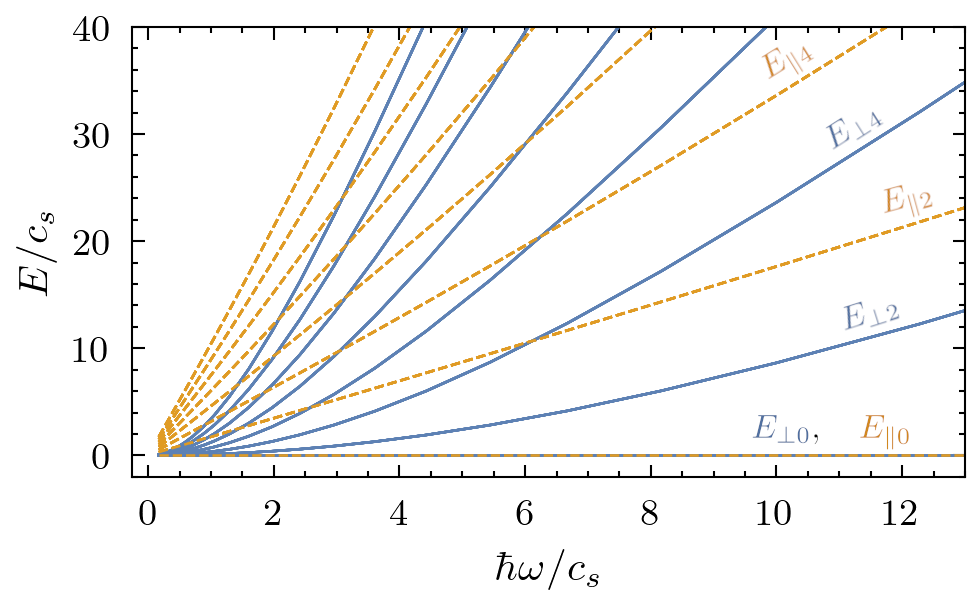}
\caption{
Bogoliubov spectrum of the Hamiltonian $\tcLscal(\vr, \vr')$ as a function of
the trap frequency $\omega$. Only eigenenergies corresponding to eigenmodes
with even spatial symmetry are shown. The blue solid and orange dashed lines
denote spin-perpendicular and spin-parallel excitations, respectively. The
value of $g_s$ is adjusted for each $\omega$ such that $c_s$ has the same value
as used in Figs.~\ref{fig:gp_ns} and \ref{fig:ffts}, namely,
$c_s = h \times 29.39$~Hz. The ratio $g_n/g_s$ is kept fixed at $28.10$ for all
calculations.
}
\label{fig:bog_spectrum}
\end{figure}

Figure~\ref{fig:bog_spectrum} shows the spectrum of the two types of
excitations as a function of the trap frequency $\omega$ for a 1D system. For
each $\omega$, the nonlinear coupling constant is adjusted such that the value
of $c_s$ is fixed. Under the Thomas-Fermi approximation, the chemical potential
$\mun = c_s (3 g_n / 5 g_s)$ is the same for all $\omega / c_s$. The blue solid
and orange dashed lines correspond to the eigenenergies of the
spin-perpendicular and spin-parallel eigenmodes, respectively. Only the
eigenmodes with even spatial symmetry are shown, as the harmonic potential and
initial state used in the numerics are both even in space. Since the
spin-perpendicular excitations have no particle-hole correlations, it is
expected that the energy depends on $L_c^{-1}$ quadratically, where $L_c$ is
the characteristic length scale of the system. Using the Thomas-Fermi
approximation, we obtain $L_c \approx \sqrt{\mun / (m \omega^2)}$, which
results in $E \propto \omega^2$. For the spin-parallel excitations,
particle-hole correlations exist and the eigenenergies depend on $L_c^{-1}$
linearly, i.e., $E \propto \omega$, matching the length-scale dependence
derived from the phonon dispersion relationship in free
space~\cite{RevModPhys.73.307}.

\subsection{Resonance conditions}
\label{sec:res_cond}

To derive explicit conditions for where spin-spatial resonances occur, we
analyze the time-dependent driving term $\tvSl(\vr, t)$. Resonant coupling to
an eigenmode of $\tcLscal(\vr, \vr')$ is triggered when the frequency of
$\tvSl(\vr, t)$ matches the corresponding eigenenergy of the eigenmode. In what
follows, we restrict ourselves to the ``far Zeeman regime," i.e., the regime
where the quadratic Zeeman shift dominates ($\abs{q} \gg c_s$). In this regime,
according to Eqs.~\eqref{eq:spin_sma} and~\eqref{eq:spin_sma_h}, we approximate
the time evolution operator in Eq.~\eqref{eq:U_SMA} by
\begin{equation}
\label{eq:sma_evolution_approx}
\vUsma(t)
\approx \exp\left( - \ic
\int_0^t \left[ q \vF_z^2 - \mus(t') \right] dt' / \hbar \right).
\end{equation}
Using this approximation, we find
\begin{multline}
\tvSl(\vr, t)
\approx \left[ g_s |\phi_0(\vr)|^2 - c_s \right]\Bigg\{
\begin{bmatrix} \cM(t) \phi_0(\vr) \vs_0(0) \\
- \cM^*(t) \phi^*_0(\vr) \vs^*_0(0) \end{bmatrix}
\\
+ \begin{bmatrix} \cC \phi_0(\vr) \vs_0(0) \\
- \cC^* \phi^*_0(\vr) \vs^*_0(0) \end{bmatrix}\Bigg\},
\end{multline}
where the matrix $\cM(t)$ has the matrix elements
\begin{equation}
\begin{aligned}
\left[\cM(t)\right]_{\alpha \beta} =
& \sqrt{2} \left[\vF_x\right]_{\alpha \beta}
\left[ \vs_0(0) \right]_{4 - \beta} \left[\vs^*_0(0)\right]_{4 - \alpha}
\cos\left(\frac{2 q t}{\hbar}\right)
-
\\
& \sqrt{2} \left[\vF_y\right]_{\alpha \beta}
\left[\vs_0(0)\right]_{4 - \beta} \left[\vs^*_0(0)\right]_{4 - \alpha}
\sin\left(\frac{2 q t}{\hbar}\right)
\end{aligned}
\end{equation}
and the time-independent matrix $\cC$ is given by
\begin{align}
\cC = \sum_{j=x,y,z} \left[\vs_0^\dagger(0) \vF_j \vs_0(0) \right] \vF_j
- \cM(0).
\end{align}
This analysis shows that the matrix elements of $\tvSl(\vr, t)$ contain terms
that oscillate at the frequency $2q/\hbar$, resulting in the resonance
conditions
\begin{equation}
\label{eq:res_fundamental_perp}
2 \abs*{q_{\text{res}}} = E_{\perp k}
\end{equation}
and
\begin{equation}
\label{eq:res_fundamental_para}
2 \abs*{q_{\text{res}}} = E_{\parallel k},
\end{equation}
where $E_{\perp k}$ and $E_{\parallel k}$ are the eigenenergies defined in
Eqs.~\eqref{eq:eigenval_perp} and~\eqref{eq:eigenval_para}, respectively.
Similar resonance conditions have also been discussed in other systems,
such as binary BECs with interaction effectively modulated by Rabi
oscillations~\cite{PhysRevA.100.063610}.

Our derivation relies on three approximations: i)~terms that are quadratic or
of higher order in $\vl(\vr, t)$ are dropped, ii)~linear terms that are
proportional to $g_s \vl(\vr, t)$ are also neglected since they are effectively
counted as being of quadratic order (see our discussion above), and iii)~the
resonance occurs in the large quadratic Zeeman shift regime, $\abs{q} \gg c_s$.
Condition i) holds until the excitation modes have a significant fraction of
the population. This can occur if the system is exactly on resonance over very
long time scales. Importantly, signatures of the resonance should emerge before
condition i) is violated. Condition ii) is almost always satisfied when
condition i) is satisfied provided $g_s$ is much smaller than the other energy
scales of the system. Recall that this scale separation is a critical
ingredient of our theory. Condition iii) tends to be satisfied in the
reasonably strong trap regime, where the energy spacing of the Bogoliubov
spectrum is much larger than the energy scale $c_s$ (see below for additional
discussion).

The neglected nonlinear terms in $\vl(\vr, t)$ in the equations of motion for
$\vl(\vr, t)$ and $\vs_0(t)$ introduce frequencies that are integer multiples
of the fundamental frequency [see Eqs.~\eqref{eq:res_fundamental_perp}
and~\eqref{eq:res_fundamental_para}]. Therefore, the more general resonance
conditions are
\begin{equation}
\label{eq:res_parametric}
2 n \abs*{q_\text{res}} = E_{\perp k}
, \quad
2 n \abs*{q_\text{res}} = E_{\parallel k}
, \quad
(n = 1, 2, 3, \dots)
\end{equation}
where $n = 1$ corresponds to the fundamental resonance frequencies derived
above, while $n > 1$ corresponds to higher-order parametric resonance
frequencies. Figure~\ref{fig:gp_ns}(b), e.g., corresponds to a parametric
resonance with $n = 2$.

The analysis presented in the previous section shows that the resonances
associated with the $\modeppar{k > 0}(\vr)$ and $\modehpar{k > 0}(\vr)$
particle-hole-correlated spatial wavefunctions correspond to a spin structure
$\vs_0(t)$ that is parallel to the background spin state, whereas those
associated with the $\modeperp{k > 0}(\vr)$ spatial wavefunctions correspond to
a spin structure $\vs_{k > 0}(t)$ that is perpendicular to the background spin
state. The former implies that different spin components share the same spatial
wavefunction to leading order near the resonances and that the resulting
dynamics resemble an overall density oscillation, whereas the latter implies
that the resonances involve changing spin structures, leading to spin-dependent
spatial wavefunctions. We use the state fidelity
\begin{equation}
\mathcal{F}_{\pm}(t)
= \frac{\abs*{ \int d\vr\, \psi^*_0(\vr, t) \, \psi_{\pm1}(\vr, t) }}
{\sqrt{\rho_0(t)\rho_{\pm 1}(t)}}
\end{equation}
to identify resonances that are associated with the spin-perpendicular modes
and the integrated time difference in the total single-particle density
\begin{equation}
\mathcal{D}(t) = \frac{1}{2} \int d\vr\,
\left| n_\text{tot}(\vr, t) - n_\text{tot}(\vr, 0) \right|
\end{equation}
to identify resonances that are associated with the particle-hole-coupled modes
(the factor of $1/2$ ensures a maximum value of unity).

\begin{figure*}[htbp]
\centering
\includegraphics[width=\textwidth]{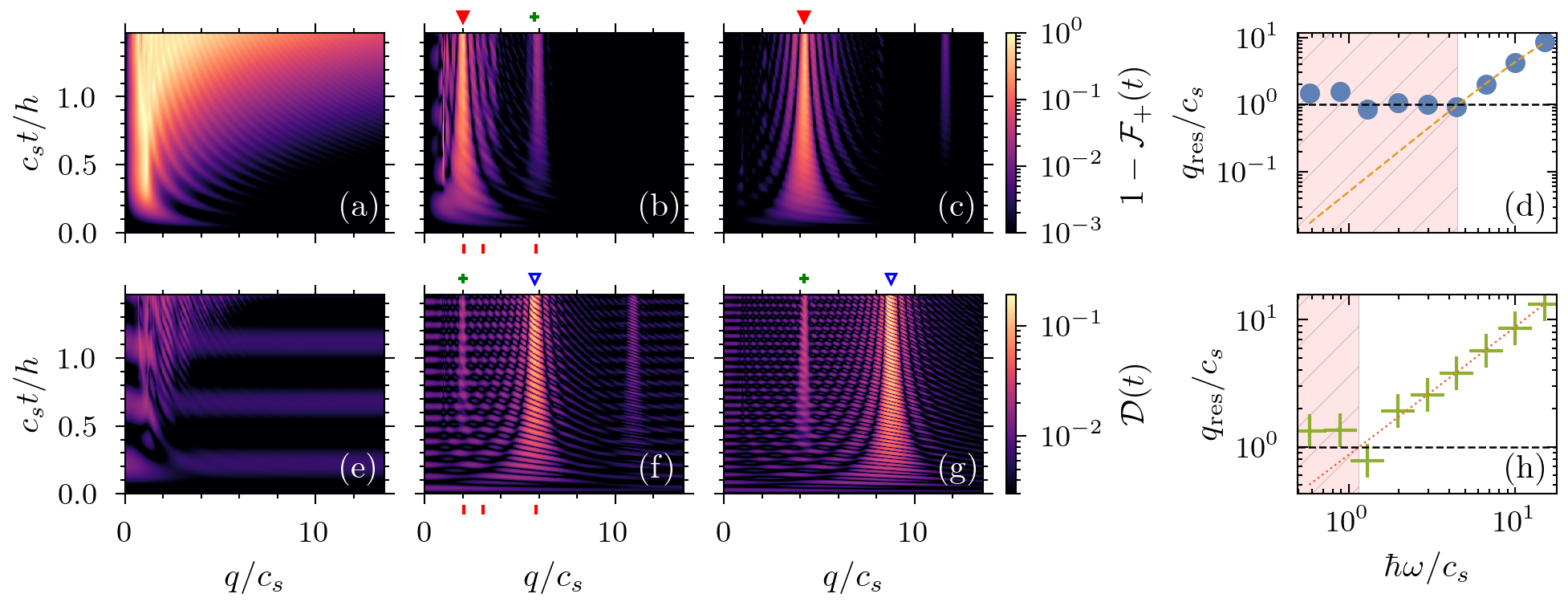}
\caption{
Resonance positions as a function of the trap frequency $\omega$ obtained from
SGPE simulations. The first, second, and third columns correspond to $\omega =
2 \pi \times 38$~Hz, $2 \pi \times 196$~Hz, and $2 \pi \times 294$~Hz,
respectively. Panels (a)–(c) and (e)–(g) show the time traces of
$1-\mathcal{F_+}(t)$, signaling spin-perpendicular resonances, and
$\mathcal{D}(t)$, signaling spin-parallel resonances, respectively, as a
function of $q$. The red lines below (b) and (f) indicate the vertical cuts
corresponding to the SGPE simulation results shown in Fig.~\ref{fig:gp_ns}. The
solid red triangles above (b) and (c) indicate resonance peaks corresponding to
the spin-perpendicular excitations, while the blue hollow triangles above (f)
and (g) indicate resonance peaks corresponding to spin-parallel excitations.
The green crosses above (b), (f), and (g) mark spin-parallel,
spin-perpendicular, and spin-perpendicular excitations, respectively. Panels
(d) and (h) compare the SGPE resonance positions extracted from the peak maxima
(blue dots and green crosses) with the Bogoliubov theory predictions [the
orange dashed and red dotted lines show
$q_\text{res}(\omega) = E_{\perp 2}(\omega) / 2$ and
$q_\text{res}(\omega) = E_{\parallel 2}(\omega) / 2$ for the spin-perpendicular
excitations and spin-parallel excitations, respectively]. The pink-shaded
regions in panels (d) and (h) indicate the “weak-trap” regimes in which no
sharp resonance positions can be identified. Inside the weak-trap regimes for
each resonance type, the maxima align with the SMA spin-separatrix value of
$q/c_s = 1$ (the black horizontal dashed line). Outside of the weak-trap
regime, the resonance positions extracted from the SGPE results show excellent
agreement with the Bogoliubov theory predictions.
}
\label{fig:q_scan}
\end{figure*}

Figure~\ref{fig:q_scan} shows the time traces of $1-\mathcal{F}_{+}(t)$ and
$\mathcal{D}(t)$ as functions of $q$ for three different trap frequencies and a
fixed value of $c_s = h \times 29.39~\text{Hz}$ for a 1D system. The resonance
positions can be extracted by locating the maxima of $1-\mathcal{F}_{+}(t)$ and
$\mathcal{D}(t)$ depending on the nature of the resonances. Within the SMA, the
dynamics changes, in classical phase space, from bound motion to unbound motion
at $q / c_s = 1$~\cite{You_PhysRevA2005}. Different features can be seen around
the SMA ``spin-separatrix value'' of $q / c_s = 1$ for all three trap
frequencies. Figures~\ref{fig:q_scan}(a)-\ref{fig:q_scan}(c) and
\ref{fig:q_scan}(e) show peaks, whose widths and intensities vary with the trap
frequency, around $q/c_s = 1$ [note the peak for (c) is very faint]. These
peaks indicate spatial excitations near the spin separatrix. In
Figs.~\ref{fig:q_scan}(f) and \ref{fig:q_scan}(g), the role of $q / c_s = 1$ as
a separatrix between two distinct regimes of dynamics is more obvious: for
$q/c_s < 1$, the striations in $\mathcal{D}(t)$ due to the breathing mode
frequency are regular and roughly independent of $q$, while for $q/c_s > 1$,
the striations become much more intricate and $q$-dependent.

The weak-trap regimes associated with the spin-perpendicular resonances and the
spin-parallel resonances shown in Figs.~\ref{fig:q_scan}(a) and
\ref{fig:q_scan}(e), respectively, are characterized by broad features for
$q/c_s \gtrsim 1$, which lack signatures of isolated resonances; isolated
resonances would appear as vertical peaks. For these parameters, the respective
Bogoliubov spectra are comparatively dense (see Fig.~\ref{fig:bog_spectrum}),
leading to overlapping resonances that cannot be individually resolved due to
the excitation of multiple modes even for short times. In contrast, in the
strong-trap regimes associated with the spin-perpendicular and spin-parallel
excitations, the spin-perpendicular Bogoliubov spectrum and spin-parallel
Bogoliubov spectrum, respectively, are comparatively sparse. In these cases,
the resonance widths are smaller, allowing for a clear identification of
individual resonances that closely match the $2 q_\text{res} = E_{\perp 2}$ and
$2 q_\text{res} = E_{\parallel 2}$ predictions [see the orange dashed and red
dotted lines in Figs.~\ref{fig:q_scan}(d) and \ref{fig:q_scan}(h),
respectively]. As mentioned at the end of Sec.~\ref{sec:bog_modes} and in
Fig.~\ref{fig:bog_spectrum}, the solutions to Eqs.~\eqref{eq:eigenval_perp}
and~\eqref{eq:eigenval_para} with odd spatial symmetry cannot be excited given
our potential and initial state and the lowest-lying resonances therefore
correspond to the first \emph{even} excitations (hence the subscript ``2" in
$E_{\perp 2}$ and $E_{\parallel 2}$). Figures~\ref{fig:q_scan}(d) and
\ref{fig:q_scan}(h) highlight the division between the weak- and strong-trap
regimes by plotting the resonance positions of the energetically lowest-lying
spin-perpendicular resonance and spin-parallel resonance extracted from the
SGPE results against the trap frequency. The numerically determined resonance
positions show a kink-like turnover, with the turnover from
$q_\text{res} / c_s$ being constant to $q_\text{res} / c_s$ being proportional
to a power of $\hbar \omega /c_s$ occurring at different values of $\hbar
\omega / c_s$ for the spin-parallel and spin-perpendicular resonances. The
turnover occurs roughly at the trap frequency for which the lowest Bogoliubov
excitation of the spectra has the energy of either $E_{\perp 2} = 2 c_s$ or
$E_{\parallel 2} = 2 c_s$ (see Fig.~\ref{fig:bog_spectrum}); these frequencies
are then used to define the boundary between the two regimes for the two types
of resonances. This implies in the strong-trap regime for the spin-parallel
resonances and the spin-perpendicular resonances, $q_\text{res}$ scales with
the first and second powers of $\hbar \omega / c_s$, respectively, as shown in
Fig.~\ref{fig:bog_spectrum}. In Figs.~\ref{fig:q_scan}(d) and
\ref{fig:q_scan}(h), the trap frequency that delineates the two regimes for
each excitation type can be identified by the frequency at which the orange
dashed line or red dotted line, which mark $q = E_{\perp 2} / 2$ and
$q = E_{\parallel 2} / 2$, respectively, crosses the horizontal black dashed
line, which marks $q / c_s = 1$.

In the strong-trap regime shown in
Figs.~\ref{fig:q_scan}(b)-\ref{fig:q_scan}(c) and~\ref{fig:q_scan}(f) and
\ref{fig:q_scan}(g), several features that are common to $1-\mathcal{F}_{+}(t)$
and $\mathcal{D}(t)$ can be identified. The widths in $q/c_s$ of the strongest
resonance peaks (marked by triangles) decrease with increasing $t$. The
strongest peak is accompanied by less intense ``side peaks,'' whose position
changes with $q/c_s$ with increasing $t$. Since the main peak is very
pronounced at later times, the resonance positions can be extracted accurately.
The resulting resonance positions show good agreement with our Bogoliubov
theory based predictions [see triangles at the top of
Figs.~\ref{fig:q_scan}(b), \ref{fig:q_scan}(c), \ref{fig:q_scan}(f), and
\ref{fig:q_scan}(g)]. It can be seen that the measure $1-{\cal{F}}_+(t)$ shows
a faint signature of the spin-parallel excitation [green plus at the top of
Fig.~\ref{fig:q_scan}(b)] and that, likewise, the measure ${\cal{D}}(t)$ shows
a faint signature of the spin-perpendicular excitation [green plus at the top
of Fig.~\ref{fig:q_scan}(f)]. Performing simulations for various trapping
frequencies, the circles and pluses in Figs.~\ref{fig:q_scan}(d) and
\ref{fig:q_scan}(h) extracted from our SGPE simulations show good agreement
with the predictions of the Bogoliubov theory for both types of resonances. We
note that Fig.~\ref{fig:q_scan}(f) also shows a somewhat fainter isolated
resonance at $q/c_s \approx 11$, which corresponds to the second-even-excited
spin-parallel mode with $E_{\parallel 4}/c_s \approx 22$ (see the energy of the
spin-parallel resonances with $\hbar \omega / c_s \approx 6.7$ in
Fig.~\ref{fig:bog_spectrum}).

\begin{figure}
\centering
\includegraphics[width=\linewidth]{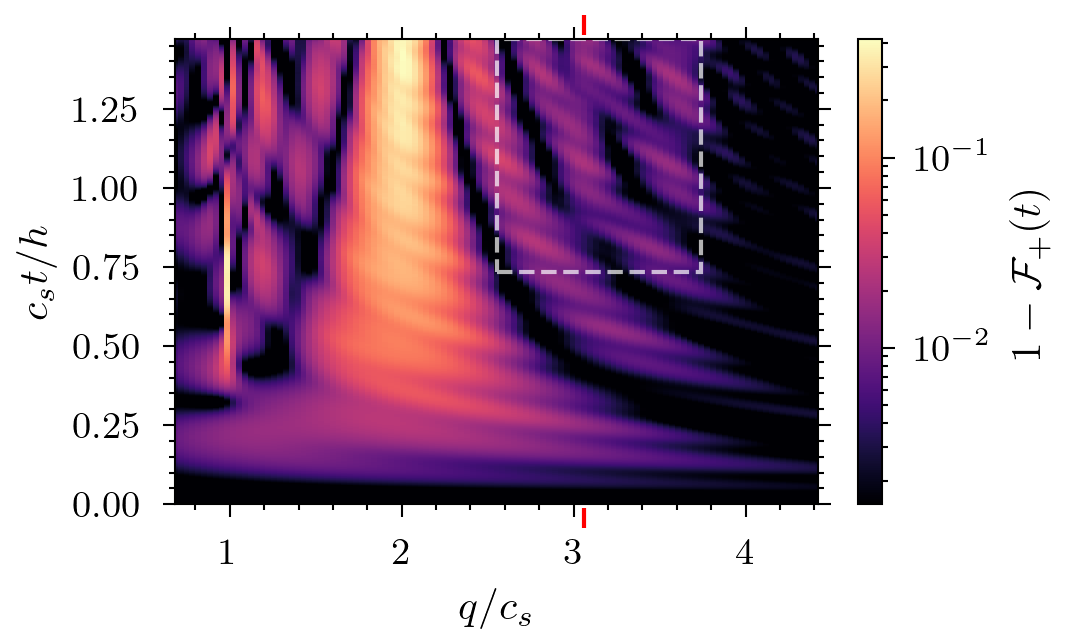}
\caption{
Zoomed view of Fig.~\ref{fig:q_scan}(b) showing $1 - \mathcal{F}_+(t)$ in the
region around the $q / c_s \approx 3$ resonance. The red vertical lines above
and below the plot show the location of the resonance predicted from the
$E_{\perp 4} = 4 q_\text{res}$ resonance condition for
$\hbar \omega / c_s = 6.668$. The white dashed box encloses the region where
the state fidelity map shows indications of a crossing between the side peaks
and a vertical feature due to the $E_{\perp 4} = 4 q_\text{res}$ resonance.
}
\label{fig:fq_zoom}
\end{figure}

The red vertical marks below Figs.~\ref{fig:q_scan}(b) and \ref{fig:q_scan}(f)
indicate the $q / c_s$ values used in Fig.~\ref{fig:gp_ns}. The marker in the
middle corresponds to the $q / c_s \approx 3$ run, which is shown in the middle
column of Fig.~\ref{fig:gp_ns}. This resonance is a parametric excitation of
the second even spin-perpendicular type, with resonance condition
$E_{\perp 4} = 4 q_\text{res}$. Figure~\ref{fig:fq_zoom} shows a zoomed and
rescaled view of Fig.~\ref{fig:q_scan}(b). The red vertical marks in
Fig.~\ref{fig:fq_zoom} indicate the location on the $x$-axis for which
$q = E_{\perp 4} / 4$. Unlike for several of the other resonances, this
resonance has a particularly low amplitude, and thus no bright discernible peak
is visible for $q / c_s \approx 3$ in Fig.~\ref{fig:fq_zoom}; instead, the
resonance manifests as crossings with the side peaks that surround the larger
nearby resonance at $q / c_s \approx 2$. The area enclosed by the white dashed
box in Fig.~\ref{fig:fq_zoom} shows the region where the crossings occur.

\begin{figure}[htbp]
\centering
\includegraphics[width=\linewidth]{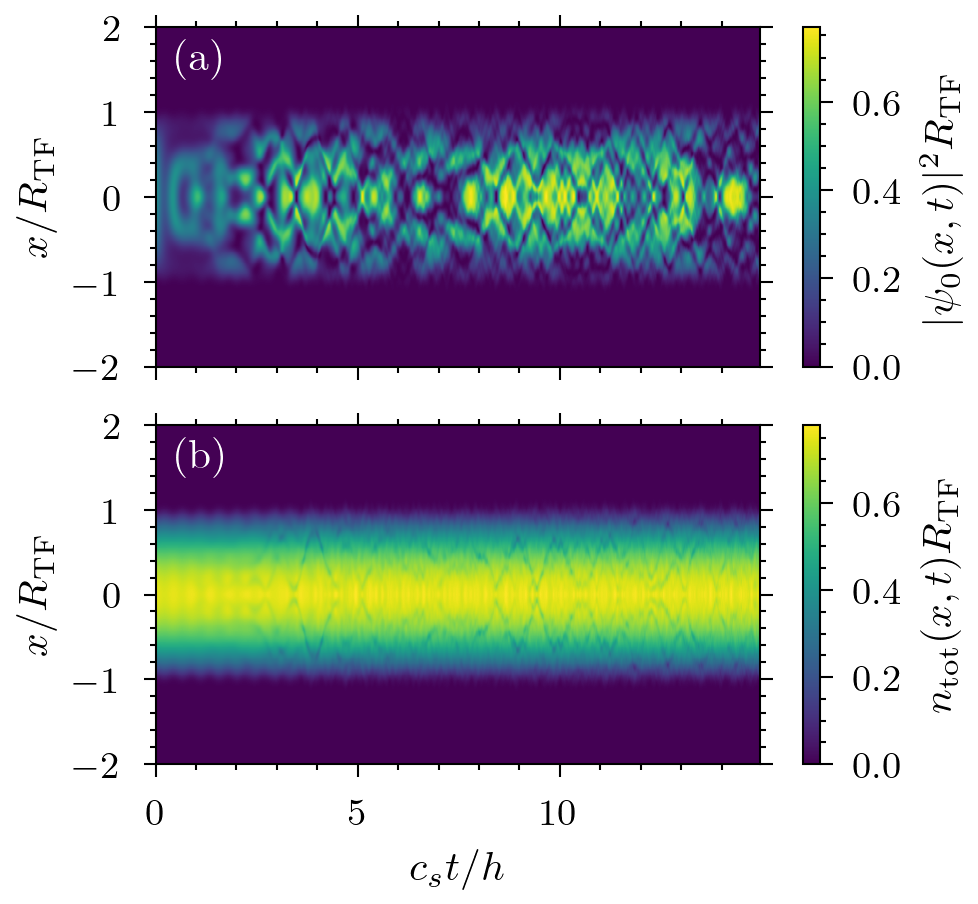}
\caption{
Spatial dynamics obtained from the SGPE in the weak-trap regime with
$\omega = 2 \pi \times 38$~Hz and $q = h \times 15$~Hz ($q / c_s = 0.512$).
Panel (a) shows the time evolution of the $m_F = 0$ component density
$|\psi_0(x, t)|^2$. Panel (b) shows the time evolution of the total condensate
density $n_\text{tot}(x, t) = \vpsi^\dagger(x, t) \vpsi(x, t)$. Despite strong
dynamics in the individual spin components, the total density is approximately
stationary throughout the evolution.
}
\label{fig:weaktrap}
\end{figure}

In the weak-trap regime the total density depends only weakly on time, as
indicated by the comparatively small values of $\mathcal{D}(t)$ in
Fig.~\ref{fig:q_scan}(e). The individual spin components, however, show
appreciable spatial dynamics, as indicated by the comparatively large values of
$1-\mathcal{F}_{+}(t)$ in Fig.~\ref{fig:q_scan}(a). This can be understood as a
partial consequence of the small ratio of the spin characteristic length
scales, i.e., the spin healing length and the length scale associated with $q$,
to the system size in the weak-trap regime, which allows for spin-domain
formation even when the total density distribution remains relatively
unchanged. Figure~\ref{fig:weaktrap} shows an example of the formation of spin
domains in the weak-trap regime. The simulation parameters are the same as
those used in the first column of Fig.~\ref{fig:q_scan} with
$q/c_s \approx 0.511$. Much later times are shown in Fig.~\ref{fig:weaktrap}
than in Fig.~\ref{fig:q_scan}. It can be seen that the $m_F = 0$ density
[Fig.~\ref{fig:weaktrap}(a)] possesses, for $c_s t / h \gtrsim 3$, ``sharp''
maxima and minima, many of which do not appear in the total density
[Fig.~\ref{fig:weaktrap}(b)]. Specifically, the total density remains roughly
unchanged throughout the dynamics, aside from some traveling deformations and
small breathing oscillations (which are faint). We interpret the long-time
dynamics as corresponding to the formation of spin domains.

\subsection{Multiple-mode expansion of the
\texorpdfstring{$\vl$}{Lambda}-model}
\label{sec:fsm}

To extend the commonly used SMA, we now use the spatial wavefunctions that form
the eigenbases of $\tcLperp(\vr, \vr')$ and $\tcLpar(\vr, \vr')$ shown in
Eqs.~\eqref{eq:eigenval_perp} and~\eqref{eq:eigenval_para} to construct a
multiple-mode expansion of the perturbation $\vl(\vr, t)$ in the unrotated
frame. By direct substitution into the unrotated equivalents of
Eqs.~\eqref{eq:lambda_spin_decomp}-\eqref{eq:bogoliubov_eigenmode_para}, we
obtain
\begin{align}
\label{eq:lambda_spin_decomp_unrotated}
\begin{bmatrix} \vl(\vr, t) \\ \vl^*(\vr, t) \end{bmatrix}
= \begin{bmatrix} \vlp(\vr, t) \\ \vlp^*(\vr, t) \end{bmatrix}
+ \begin{bmatrix} \vls(\vr, t) \\ \vls^*(\vr, t) \end{bmatrix}
\end{align}
with
\begin{align}
\label{eq:vl_w_modes_perp}
\begin{bmatrix} \vlp(\vr, t) \\ \vlp^*(\vr, t) \end{bmatrix}
= \sum_{k>0} \begin{bmatrix} \vs_k(t) \modeperp{k}(\vr) \\ 0 \end{bmatrix}
+ \begin{bmatrix} 0 \\ \vs_k^*(t) \modeperp{k}^*(\vr) \end{bmatrix}
\end{align}
and
\begin{align}
\label{eq:vl_w_modes_para}
&\begin{bmatrix} \vls(\vr, t) \\ \vls^*(\vr, t) \end{bmatrix}
\\ \nonumber
&= \sum_{k > 0} \left( b_k(t) \begin{bmatrix}
\vs_0(t) \modeppar{k}(\vr)
\\
\vs_0^*(t) \modehpar{k}(\vr) \end{bmatrix}
+ b_k^*(t) \begin{bmatrix}
\vs_0(t) \modehpar{k}^*(\vr)
\\
\vs_0^*(t) \modeppar{k}^*(\vr) \end{bmatrix}
\right),
\end{align}
where we have
\begin{multline}
\cLscal(\vr, \vr', t)
\begin{bmatrix} \vl(\vr', t) \\ \vl^*(\vr', t) \end{bmatrix}
= \sum_{k>0} E_{\perp k} \begin{bmatrix} \vs_k(t) \modeperp{k}(\vr)
\\ - \vs_k^*(t) \modeperp{k}^*(\vr) \end{bmatrix}
\\
+ \sum_{k > 0} E_{\parallel k} \left(
b_k(t) \begin{bmatrix} \vs_0(t) \modeppar{k}(\vr)
\\
\vs_0^*(t) \modehpar{k}(\vr) \end{bmatrix}
- b_k^*(t) \begin{bmatrix} \vs_0(t) \modehpar{k}^*(\vr)
\\
\vs_0^*(t) \modeppar{k}^*(\vr) \end{bmatrix}
\right).
\end{multline}
The terms on the right hand side of Eqs.~\eqref{eq:vl_w_modes_perp}
and~\eqref{eq:vl_w_modes_para} represent the dominant excitations of the
system. These modes provide the dominant structure of the excitations of the
system as suggested by Eq.~\eqref{eq:eom_hl_lgp_rotated_approx}. The neglected
terms not present in Eq.~\eqref{eq:eom_hl_lgp_rotated_approx} include
higher-order terms and the action of the spin-part $\cLs(\vr, \vr, t)$ of the
full Bogoliubov matrix $\cLgp(\vr, \vr', t)$ on the perturbation, which
together provide nonlinearities and mode couplings.

Using othonormality of the basis states, the spin structure associated with the
$k$-th spin-perpendicular Bogoliubov mode can be written as
\begin{equation}
\vs_k(t) = \cQs(t) \int d\vr \modeperp{k}^*(\vr) \vl(\vr, t), \qquad k > 0
\end{equation}
and the mode occupation for the $k$-th spin-parallel Bogoliubov mode as
\begin{multline}
b_k(t) = \int \modeppar{k}^*(\vr) \vs_0^\dagger(t) \vl(\vr, t) d\vr
\\
- \int \vl^\dagger(\vr, t) \vs_0(t) \modehpar{k}^*(\vr) d\vr, \qquad k > 0.
\end{multline}
Using the last two equations, we can obtain the corresponding equations of
motion by taking time derivatives, remembering that the operators are time
dependent and the spatial wavefunctions are time independent. We find
\begin{multline}
\label{eq:eom_sperp}
\ic \hbar \frac{\partial}{\partial t} \vs_k(t) =
\\
\begin{aligned}
&\vs_0(t)
\left[ \ic \hbar \frac{\partial}{\partial t} \vs_0(t) \right]^\dagger
\vs_k(t)
\\
&\begin{aligned}
+ \int d\vr \modeperp{k}^*(\vr) &\left[
\cQs(t) \left[ \ic \hbar \frac{\partial}{\partial t} \vl(\vr, t) \right]
\right.
\\
&\ \left.
- \left[ \ic \hbar \frac{\partial}{\partial t} \vs_0(t) \right]
\vs_0^\dagger(t) \vls(\vr, t) \right],
\end{aligned}
\end{aligned}
\\
k > 0,
\end{multline}
and
\begin{multline}
\label{eq:eom_b}
\ic \hbar \frac{\partial}{\partial t} b_k(t) =
\\
\begin{aligned}
\int d\vr \modeppar{k}^*(\vr) & \left[ \vs_0^\dagger(t)
\left[ \ic \hbar \frac{\partial}{\partial t} \vl(\vr, t) \right] \right.
\\
&\ \left.
- \left[ \ic \hbar \frac{\partial}{\partial t} \vs_0(t) \right]^\dagger
\vlp(\vr, t) \right]
\end{aligned}
\\
\begin{aligned}
+ \int d\vr & \left( \left[
\ic \hbar \frac{\partial}{\partial t} \vl(\vr, t) \right]^\dagger \vs_0(t)
\right.
\\
&\ \ \ \left.
- \vlp^\dagger(\vr, t) \left[
\ic \hbar \frac{\partial}{\partial t} \vs_0(t) \right] \right)
\modehpar{k}^*(\vr),
\end{aligned}
\\
k > 0.
\end{multline}
Equations~\eqref{eq:eom_sperp} and~\eqref{eq:eom_b} show that the parallel and
perpendicular subspaces are coupled via spin dynamics; this coupling arises
from the non-orthogonality of the two subspaces [see
Eqs.~\eqref{eq:mode_overlap_p} and~\eqref{eq:mode_overlap_h}]. While each mode
of one subspace is, in principle, coupled to all modes of the other via spin
dynamics, we find in practice that modes of significantly different energies
are only very weakly coupled. To obtain closed-form expressions for the
equations of motion for $\vs_k(t)$ and $b_k(t)$, one must: replace the time
derivatives within the square brackets in Eqs.~\eqref{eq:eom_sperp}
and~\eqref{eq:eom_b}, namely $\ic \hbar \partial \vs_0(t) / \partial t$ and
$\ic \hbar \partial \vl(\vr, t) / \partial t$ and their adjoints, using the
right hand sides of Eqs.~\eqref{eq:eom_s0} and \eqref{eq:eom_lambda} and their
adjoints; replace $N_0(t)$, which enters as part of the elimination, by the
right hand side of Eq.~\eqref{eq:N0}; and then lastly replace the remaining
$\vl(\vr, t)$ by the right hand side of
Eqs.~\eqref{eq:lambda_spin_decomp_unrotated}-\eqref{eq:vl_w_modes_para}. Doing
so yields a closed set of equations, namely we obtain
$\partial \vs_k(t) / \partial t$ and $\partial b_{k > 0}(t) / \partial t$ as
functions of $\vs_k(t)$, $b_{k > 0}(t)$, and time-independent quantities. As
one might imagine, the full expressions contain a multitude of terms and are
not particularly illuminating. In what follows, we focus on a truncated version
of these expressions and analyze an unrotated ``two-state" equivalent of the
truncated model described in Sec.~\ref{sec:bog_modes}. We clarify that when we
count the number of states included in this model, we explicitly count the
number of unique mode amplitudes included, i.e., the number of $\vs_k(t)$
(including $k = 0$) and $b_k(t)$ terms but not their conjugates, which are
implicitly assumed to be included in the expansion as well.

Given that a description that only includes the terms shown in
Eqs.~\eqref{eq:eom_s0_expansion} and~\eqref{eq:eom_vl_lgp} already abandons,
due to the neglected quadratic terms in $\vl(\vr, t)$ in
Eq.~\eqref{eq:eom_s0_expansion}, Hermiticity and therefore particle number
conservation, we assume $N_0(t) = 1$ in what follows, i.e., we assume that the
background serves as a particle reservoir, which seeds the dynamics of the
resonant modes. We truncate the equations of motion based on the two-body
interaction integrals that appear. Explicitly, in the following models we
assume that any two-body interaction integral proportional to $g_n$ involving
less than two combined powers of $\phi_0(\vr)$ and $\phi^*_0(\vr)$ is
negligible, and any two-body interaction integral proportional to $g_s$
involving less than three combined powers of $\phi_0(\vr)$ and $\phi^*_0(\vr)$
is negligible. Retaining nonlinear terms in the resonant mode amplitudes in
accordance with these assumptions ensures that
$\vs_0^\dagger(t) \vs_{k > 0}(t)$ identically vanishes. Depending on the type
of resonance, the resonant modes are those with the particle-components given
by $\vs_2(t) \modeperp{2}(\vr)$ or
$\vs_0(t) [ b_2(t) \modeppar{2}(\vr) + b_2^*(t) \modehpar{2}^*(\vr) ]$. If the
resonant wavefunction is $\vs_2(t) \modeperp{2}(\vr)$, we obtain for the
background spin
\begin{multline}\label{eq:eom_s0_2perp}
\ic \hbar\frac{\partial \vs_0(t)}{\partial t} \approx
\\
\left[ \vhs(t) - \mus(t) \right] \vs_0(t) + g_s \cQs(t) \cPperp(t)
\end{multline}
and for the resonant spinor amplitude
\begin{align}\label{eq:eom_s2_2perp}
\ic \hbar \frac{\partial \vs_2(t)}{\partial t}
\approx{}& g_s \alpha \cQs(t) \left[ \sum_{j=x,y,z}
\vs_0^\dagger(t) \vF_j \vs_0(t) \vF_j \right] \vs_0(t)
\nonumber \\
& + \left[ q \cQs(t) \vF_z^2 + E_{\perp 2} \right] \vs_2(t)
\nonumber \\
& - \left[ \mus(t) + g_s \cPperp^\dagger(t) \vs_0(t) \right] \vs_2(t),
\nonumber \\
& + \vs_0(t)
\left[ \ic \hbar \frac{\partial}{\partial t} \vs_0(t) \right]^\dagger
\vs_2(t)
\end{align}
where
\begin{equation}
\begin{aligned}[b]
\cPperp(t) = \sum_{j=x,y,z}
& \alpha^* \vs_0^\dagger(t) \vF_j \vs_0(t) \vF_j \vs_2(t)
\\
+ & \alpha^* \vF_j \vs_0(t) \vs_0^\dagger(t) \vF_j \vs_2(t)
\\
+ & \alpha \vF_j \vs_0(t) \vs_0^\trans(t) \vF_j^\trans \vs^*_2(t)
\end{aligned}
\end{equation}
and
\begin{equation}
\alpha = \int d\vr \modeperp{2}^*(\vr) \abs*{\phi_0(\vr)}^2 \phi_0(\vr).
\end{equation}
If the resonant wavefunction is
$\vs_0(t) [ b_2(t) \modeppar{2}(\vr) +b_2^*(t) \modehpar{2}^*(\vr) ]$, we
obtain
\begin{align}\label{eq:eom_s0_2par}
\ic \hbar \frac{\partial \vs_0(t)}{\partial t}
\approx{}& \left[\vhs(t) - \mus(t)\right]\vs_0(t)
\nonumber \\
& + g_s \cQs(t) \cPpar(t)
\end{align}
and
\begin{multline}\label{eq:eom_b2_2par}
\ic \hbar \frac{\partial b_2(t)}{\partial t}
\approx g_s (\beta + \gamma^*)
\sum_{j=x,y,z} \left[ \vs_0^\dagger(t) \vF_j \vs_0(t) \right]^2
\\
+ \left\{ E_{\parallel 2}
- \ic g_s \Im\left[ \cPpar^\dagger(t) \vs_0(t) \right]
\right\} b_2(t)
\\
+ \left[ q \vs_0^\dagger(t) \vF_z^2 \vs_0(t) - \mus(t)
- g_s \Re\left[ \cPpar^\dagger(t) \vs_0(t) \right]
\right] \times
\\
\left[ \zeta_0 b_2(t) + \zeta_1 b_2^*(t) \right],
\end{multline}
where
\begin{align}
\cPpar(t) ={}& \left[ (2 \beta^* + \gamma) b_2(t)
+ (\beta + 2 \gamma^*) b_2^*(t) \right] \times
\nonumber \\
& \MoveEqLeft[16] \left[ \vs_0^\dagger(t) \vF_j \vs_0(t) \vF_j \right]
\vs_0(t),
\\
\beta ={}& \int d\vr \modeppar{2}^*(\vr) \abs*{\phi_0(\vr)}^2 \phi_0(\vr),
\\
\gamma ={}& \int d\vr \modehpar{2}(\vr) \abs*{\phi_0(\vr)}^2 \phi_0(\vr),
\\
\zeta_0 ={}& \int d\vr \left[ \abs*{\modeppar{2}(\vr)}^2
+ \abs*{\modehpar{2}(\vr)}^2 \right],
\\
\zeta_1 ={}& 2 \int d\vr \modeppar{2}^*(\vr) \modehpar{2}^*(\vr),
\end{align}
and $\Re[\dots]$ and $\Im[\dots]$ denote the real and imaginary parts of the
argument, respectively.

The equations of motion for $\vs_2(t)$ and $b_2(t)$
[Eqs.~\eqref{eq:eom_s2_2perp} and~\eqref{eq:eom_b2_2par}] each contain a
seeding term proportional to the integral coefficients $\alpha$ and
$(\beta+\gamma^*)$, respectively. For an initial state with vanishing
$\vs_2(0)$ and $b_2(0)$, resonant excitations are absent if the corresponding
integral coefficients vanish. One example is the homogeneous system, in which
the background condensate has zero momentum while all Bogoliubov modes have
finite momentum. Consequently, $\alpha=\beta=\gamma=0$ due to the orthogonality
of different momentum eigenstates.

\begin{figure}[htbp]
\centering
\includegraphics[width=0.45\textwidth]{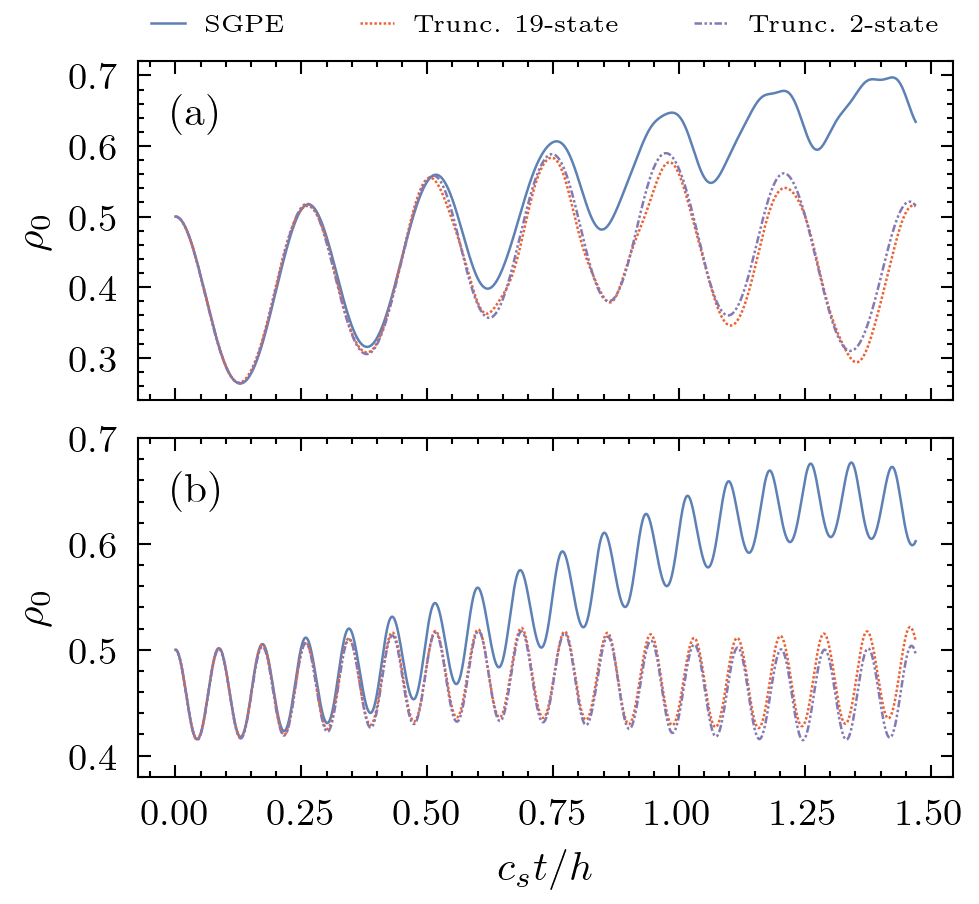}
\caption{
The dynamics of the population of the $m_F = 0$ spin component with
$\omega = 2 \pi \times 196$~Hz at two different $q$ values, (a) $q = h \times
60$~Hz and (b) $q \approx h \times 171$~Hz. The blue solid, red dotted, and
purple dash-double-dotted lines correspond to the results obtained from SGPE
simulations, the truncated $\vl$-model using 19 states with no
beyond-Bogoliubov terms, and the two-state truncated $\vl$-model with no
beyond-Bogoliubov terms, respectively. All the other parameters are the same as
the ones used in Fig.~\ref{fig:gp_ns}. The truncated models fail to capture the
SGPE dynamics accurately for $c_s t / h \gtrsim 0.8$.
}
\label{fig:rho0_trunc}
\end{figure}

Figure~\ref{fig:rho0_trunc} shows the dynamics of the $m_F=0$ population near
two resonances obtained using different few-state models and the SGPE. The
purple dash-double-dotted curves show the results from the two-state models
derived above, using Eqs.~\eqref{eq:eom_s0_2perp} and~\eqref{eq:eom_s2_2perp}
for Fig.~\ref{fig:rho0_trunc}(a) and Eqs.~\eqref{eq:eom_s0_2par}
and~\eqref{eq:eom_b2_2par} for Fig.~\ref{fig:rho0_trunc}(b). The
spin-perpendicular model (purple dash-double-dotted line) reproduces the SGPE
results (blue line) with decent agreement for $c_s t / h \lesssim 0.8$.
Importantly, both truncated models neglect beyond-Bogoliubov terms, which are
crucial for capturing the full SGPE dynamics. Specifically, the full
$\vl$-model shown in Fig.~\ref{fig:gp_ns} reproduces the SGPE results extremely
well. This indicates that the beyond-Bogoliubov terms are significant in
determining the dynamics when the system parameters are tuned closely to these
strong resonances. The neglected higher-order terms allow for transitions from
the resonant modes to even higher energy modes, modifying the chemical
potential and (to a lesser extent) the total density distribution, which in
turn may modify the resonance condition. Because $N_0(t)$ is fixed to unity in
the truncated models used in Fig.~\ref{fig:rho0_trunc}, the background state
remains undepleted, which reduces such a dynamically off-resonant effect.

Figure~\ref{fig:rho0_trunc}(b) shows results for a spin-parallel resonance
using the same line styles as used in Fig.~\ref{fig:rho0_trunc}(a). Here again,
the two-state model [now from Eqs.~\eqref{eq:eom_s0_2par}
and~\eqref{eq:eom_b2_2par}] (purple dash-double-dotted line) matches the
19-state model (red dotted line) relatively closely, with both showing some
evidence of a resonance and two-state dynamics. Importantly, though, neither of
these two models captures the resonant SGPE dynamics for
$c_s t / h \gtrsim 0.4$, indicating that density fluctuations play a
significant role in the dynamics of the spin-parallel resonance, likely due to
the parametric nature of the resonance, which is driven by large fluctuations
of the nonlinear terms.

\begin{figure}[htbp]
\centering
\includegraphics[width=0.48\textwidth]{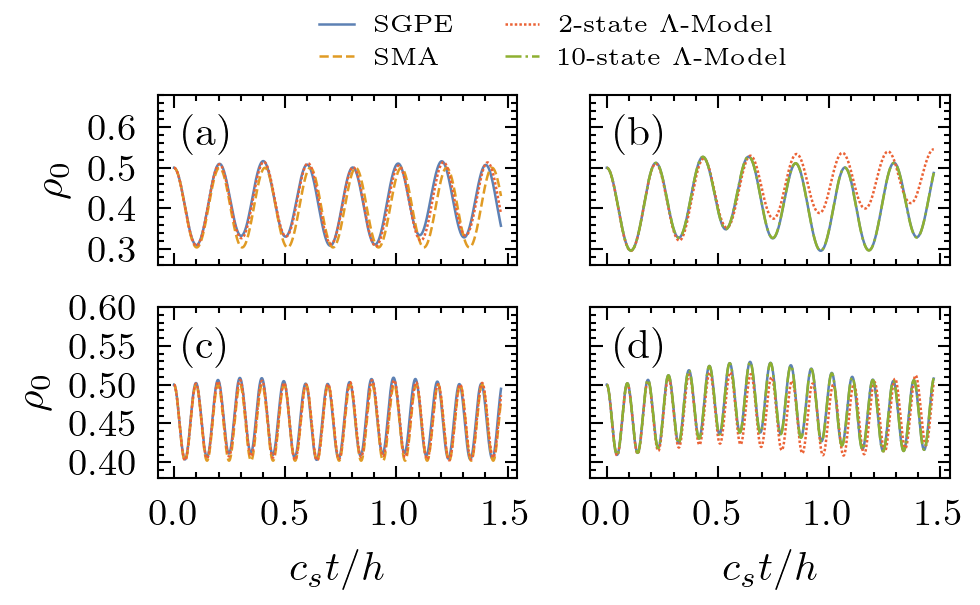}
\caption{
The dynamics of the population of the $m_F=0$ spin component with $\omega = 2
\pi \times 196$~Hz for (a-b) $q$ just above $h \times 60$~Hz [see
Fig.~\ref{fig:rho0_trunc}(a)] and (c-d) $q$ just below $h \times 171$~Hz [see
Fig.~\ref{fig:rho0_trunc}(b)]. The values of the quadratic Zeeman shifts are
(a) $q = h \times 75$~Hz, (b) $q= h \times 70$~Hz, (c) $q = h \times 150$~Hz,
and (d) $q = h \times 160$~Hz. All other parameters are the same as the ones
used in Fig.~\ref{fig:gp_ns}. In (a) and (c), the blue solid, orange dashed,
and red dotted lines correspond to the results obtained from SGPE simulations,
SMA, and $\vl$-model with all terms using two spatial wavefunctions,
respectively. The SGPE and $\vl$-models produce similar results, while the SMA
fails to capture the slow resonant oscillation. In (b) and (d), the blue solid,
red dotted, and green dash-dotted lines correspond to the results obtained from
SGPE simulations, $\vl$-model with all terms using two spatial wavefunctions,
and $\vl$-model with all terms using 10 spatial wavefunctions, respectively.
The $\vl$-model with 10 states reproduces the SGPE results, while the two-state
model fails to capture the amplitude and frequency of the slow resonant
oscillation. The slow oscillation frequencies appear to decrease in amplitude
for the values of $q$ further away from the resonant values.
}
\label{fig:rho0_detuned}
\end{figure}

The higher-order terms are kept in the models shown in
Fig.~\ref{fig:rho0_detuned}, and instead the system is probed with varying $q$
that is detuned from the resonance positions identified in
Fig.~\ref{fig:q_scan}. Results from varying mode-truncations are shown for each
$q$. The left column represents results for the $q$ values for which the system
is further detuned from the resonance peak, while the right column shows
results for the $q$ values closer to resonance.
Figures~\ref{fig:rho0_detuned}(a) and~\ref{fig:rho0_detuned}(b) detune from the
spin-perpendicular excitation at $q \approx h \times 60$ Hz, while
Figs.~\ref{fig:rho0_detuned}(c) and~\ref{fig:rho0_detuned}(d) detune from the
spin-parallel excitation at $q \approx h \times 171$ Hz. In the far detuned
cases [Figs.~\ref{fig:rho0_detuned}(a) and~\ref{fig:rho0_detuned}(c)], the
two-state model (red dotted lines) is sufficient to capture the dynamics to
good accuracy, showing slight improvements over the SMA results via
reproduction of the slower resonant oscillation frequency, seen as a shift away
from the SMA solution given by the orange dashed line with around 1 period
shown in Fig.~\ref{fig:rho0_detuned}(a) and 2 periods shown in
Fig.~\ref{fig:rho0_detuned}(c). When $q$ is closer to the resonance values
[Figs.~\ref{fig:rho0_detuned}(b) and~\ref{fig:rho0_detuned}(d)], more modes are
required to achieve similar accuracy, with 10-state model results shown as
green dash-dotted lines identically reproducing the SGPE (blue solid lines),
while the two-state models (red dotted) show disagreements about the amplitude
and frequency of the slow resonant oscillation. The agreement is especially
poor given the parameters of Fig.~\ref{fig:rho0_detuned}(b), for which the
two-state model seems to show a shifted resonance position closer to this
detuned $q$-value of 70 Hz. Note the 10 needed states in the slightly detuned
cases are fewer than the 19 states needed to reproduce the SGPE for the very
low detuning cases shown in Fig.~\ref{fig:gp_ns}, suggesting a simple model
with fewer states holds in the far-detuned parameter space while still
capturing beyond-SMA physics. The slow oscillation frequencies visible in
Fig.~\ref{fig:rho0_detuned} also appear to decrease when $q$ is closer to
resonant values, which is consistent with the discussion of the cusps shown in
Fig.~\ref{fig:q_scan}.

From the discussions above, it is clear that the nonlinear nature of the system
obfuscates the role of the various terms in the equations of motion. Despite
this, insights into the dynamics can be gained from simplified and truncated
treatments of the system. The fundamental resonance positions can be well
predicted by the Bogoliubov theory, with the quadratic Bogoliubov-like terms in
the equations of motion defining the structure of the resonance modes and the
resonant parameter regimes. The terms independent of $\vl$ in the equations of
motion for the resonant mode amplitudes seed occupation of the resonant modes
due to nonequilibrium spin dynamics, while the nonlinear terms in $\vl$ further
enhance the resonance by allowing for transitions to higher energy modes and
therefore stronger fluctuations in the mean-field interaction energies to drive
parametric excitations. The resonances can be made (artificially) weaker by
either removing the higher-energy modes, equivalently truncating the nonlinear
terms (which therefore reduces the number of modes that can be excited), or by
detuning the system from the resonant regimes such that the nonlinear terms and
higher-energy mode occupations are less significant in driving the dynamics.

\section{Conclusion and outlook}

In this work, we developed a general coupled-channel framework to describe
resonant spatial excitations induced by the spin dynamics of a spinor
Bose–Einstein condensate. Taking advantage of the scale separation, our theory
is closely related to the particle-number–conserving Bogoliubov theory for the
spin independent part of the full spinor Hamiltonian, in which the
eigenenergies and eigenstates determine the resonance conditions and the
corresponding resonant spatial wavefunctions. Depending on the nature of the
spatial excitations at resonance, two types of resonances are identified: one
involving particle–hole correlations and the other without such particle-hole
correlations. Our theory reveals that these two types of excitations are
associated with distinct spin structures: the former corresponds to a spin
structure parallel to the background spin, whereas the latter corresponds to a
spin structure perpendicular to the background spin. For resonances involving
perpendicular spin excitations, spin-dependent spatial structures emerge,
indicating a breakdown of the single-mode approximation. In contrast,
resonances involving parallel spin excitations behave more like total-density
excitations with different spin components sharing common spatial orbitals.
Owing to the nonlinear nature of the system, higher-order parametric resonances
are observed for both types of resonances.

The dynamics near resonances were studied in both the strong-trap and weak-trap
regimes. Sharp resonances could be identified only in the strong-trap regime,
where the energy spacing in the Bogoliubov spectrum is larger than the
spin-changing collision energy. In the weak-trap regime, many spatial
wavefunctions are involved, leading to more complicated spatial structures such
as the development of spin domains. The behavior in the weak-trap regime more
closely resembles that of homogeneous systems where the Bogoliubov theory of
the spin-dependent part of the full spinor Hamiltonian plays a role equally
important as that of the spin-independent part, and is known to support
dynamical instabilities in certain parameter regimes~\cite{KAWAGUCHI2012253}.
Our theory smoothly connects the sharp-resonance regime, which requires
comparatively large trapping frequencies, and the regime where dynamical
instabilities dominate, which requires comparatively small trapping
frequencies.

The framework presented in this paper focuses on the mean-field limit and
reproduces results obtained by solving the spinor Gross–Pitaevskii equation.
From a technical perspective, the developed coupled-channel framework may
enhance the efficiency of simulating higher-dimensional systems near resonances
compared with full Gross–Pitaevskii simulations, since a fairly small number of
spatial orbitals is expected to capture the resonant dynamics. In this work, we
focused---for illustrative purposes---on the resonant dynamics in a
one-dimensional system. We found that the number of orbitals required to reach
convergence depends on several factors, including the detuning from the
resonance, the relevant time scales, and the coupling strength to the resonant
modes. A detailed investigation of higher-dimensional systems is an interesting
direction for future work.

Extending our framework to include quantum fluctuations and quantum
correlations should be fairly straightforward. Such an extension would involve
deriving equations of motion for correlation functions of the excitations order
by order in terms of the second-quantized operator-equivalent of $\vl(\vr, t)$.
Extensions of our framework along these lines are closely related to the
Zaremba–Nikuni–Griffin formalism~\cite{Zaremba1999}, which is useful for
describing the dynamics of a condensate coupled to a thermal ensemble at finite
temperature. In this context, our framework can be generalized to capture both
single-particle–type excitations and excitations involving particle–hole
correlations in the thermal bath. A natural and interesting question for future
work is to investigate the robustness of the resonances against quantum and
thermal fluctuations.

Furthermore, this work opens broader avenues for using spinor BECs as quantum
simulators. For example, within the single-mode approximation, the spin
dynamics of a spin-1 system is known to be integrable~\cite{You_PhysRevA2005}.
When multiple modes become involved near resonance, the integrability of the
spin Hamiltonian is broken, leading to chaotic
dynamics~\cite{Garcia-March_2018} and enabling the exploration of richer
nonequilibrium quantum physics. By carefully engineering the coupling between
spin and spatial excitations, a trapped background condensate can mimic the
role of a cavity in which the spin is coupled to tunable cavity modes,
providing potential applications for simulating various iconic spin–photon
models. In addition, given the sensitivity of the spin dynamics near resonance
and the scale separation of the system, the spin degrees of freedom may be used
to probe quantum phenomena in spatial excitations, such as Beliaev damping
induced by phonon–phonon collisions~\cite{PHUC2013158} or Landau damping, which
arises from collisions between phonons and thermal atoms~\cite{Shchedrin2018}.

The current framework can also be applied to other cold-atom quantum
simulators, including those based on Floquet engineering, synthetic spin–orbit
coupling, or momentum-space lattices, where the pseudo-spin is encoded in
various ways and a simple frozen spatial wavefunction is typically
assumed~\cite{RevModPhys.86.153, PRXQuantum.2.017003}. The framework can be
directly used to verify the validity of these assumptions and to provide
guidance on the tunability and stability of those systems.

\section{Acknowledgements}

W. W. and Q. G. acknowledge funding support from NSF through Grant No.
PHY-2409600 and from the WSU Claire May \& William Band Distinguished
Professorship Award. D. B. acknowledges support by the Air Force Office of
Scientific Research under Award No. FA9550-24-1-0106.

\appendix
\section{Numerical details}
\label{app:numerics}

Our numerical simulations use 1D spatial grids with sizes and spacings that
vary with the trap frequency to ensure convergence of the results, meaning the
simulations do not change further with larger and finer grids. For the angular
trap frequency of $\omega = 2 \pi \times 196$~Hz (this corresponds to
$\hbar \omega = 6.668 c_s$) used in
Figs.~\ref{fig:gp_ns},~\ref{fig:ffts},~\ref{fig:fft_mode},%
~\ref{fig:rho0_trunc}, and~\ref{fig:rho0_detuned}, we use a spatial grid that
contains 256 points and is $6.114 R_\text{TF}$ (30~\textmu m) wide, i.e., $x$
goes from $- (6.114 / 2) R_\text{TF}$ to $(6.114 / 2) R_\text{TF}$, with
spin-dependent collision energy $c_s/h = 29.39$~Hz, spin-independent coupling
constant $g_n = 47.72 / (c_s R_\text{TF})$, and the coupling constant ratio
$g_n / g_s = 28.10$. Figures~\ref{fig:bog_spectrum},~\ref{fig:q_scan},
and~\ref{fig:weaktrap} use angular trapping frequencies different from the one
used in
Figs.~\ref{fig:gp_ns},~\ref{fig:ffts},~\ref{fig:fft_mode},%
~\ref{fig:rho0_trunc}, and~\ref{fig:rho0_detuned}. For
Figs.~\ref{fig:bog_spectrum}, \ref{fig:q_scan}, and \ref{fig:weaktrap}, we set
the angular trap frequency $\omega$ to the desired value and vary the
spin-dependent coupling constant $g_s$ for each $\omega$ such that $c_s/h$ is
equal to $29.39$~Hz (this value is always used). We then set $g_n$ via
$g_n = 28.10 g_s$. This implies that both $g_n$ and $g_s$ change while their
ratio is kept fixed. The initial state $\phi_0(x)$ is prepared via imaginary
time propagation using the scalar Hamiltonian [Eq.~\eqref{eq:h_scalar}], until
$\tau = - \ic t=- 100 \ic$~ms (for the $c_s$ used in this work, this
corresponds to $- 2.939 \ic h / c_s$) using the 4th-order Runge-Kutta method
(we use 8273 time steps). The full initial condensate state is then created by
multiplying the spatial wavefunction $\phi_0(x)$ by the spinor $\vs_0(0)$ [see
Eq.~\eqref{eq:sma_ansatz}]. The initial spinor $\vs_0(0)$ is chosen to be the
$m_{F_x} = +1$ state, which is far from equilibrium in the spin domain for
positive $q$ and $g_s$, as considered in this work. For the real time
evolution, we employ periodic boundary conditions and evaluate the kinetic
energy in the Hamiltonian of the SGPE via the FFT. The real time evolution of
the full SGPE Hamiltonian uses the same time step as our imaginary time
propagation.

To calculate the amplitudes $|\psi_0(x,E)|$ shown in Fig.~\ref{fig:ffts}, we
evolve the initial state under the full SGPE Hamiltonian in real time until a
final time of $t_f = 2,000$~ms. This translates into a frequency resolution of
$0.5$~Hz for the Fourier transform results shown in Fig.~\ref{fig:ffts}. The
Fourier transform of the real-time SGPE orbitals $\psi_{m_F}(x,t)$ utilizes a
Hamming window~\cite{1978IEEEP..66...51H} to mitigate artifacts, which arise
from the finite time window of our simulation results (the Fourier transform
assumes periodicity of the time signal). The Hamming or ``raised cosine" window
forces the signal to smoothly approach zero at the edges of the time window,
thereby removing the discontinuity at the edges and reducing the ``spectral
leakage'' that would occur otherwise. The Hamming window has the form
\begin{equation}
w(l) = 0.54 - 0.46 \cos\left( \frac{2 \pi l}{N_t - 1} \right),
\quad 0 \le l \le N_t - 1
\end{equation}
where $N_t$ is the total number of time samples and $l$ is the index of the
time sample. The time-domain signal is multiplied by the window element-wise
before performing the Fourier transform, which tapers the signal at the edges
and widens the peaks in the Fourier spectrum while lowering the amplitude of
the spurious values between the peaks. The application of the Hamming window
significantly improves the stability of the spinor wave functions in
energy-space (i.e., the results are---for sufficiently large
$t_f$---essentially independent of $t_{f}$).

\bibliography{main.bib}

\end{document}